\documentclass[iop]{emulateapj}

\usepackage{graphicx}
\usepackage{epstopdf}
\usepackage{natbib}
\usepackage{color}

\shorttitle{Friends of Hot Jupiters III: NIRSPEC}
\shortauthors{Piskorz et al.}

\begin{document}
\newcommand{\teff}{$T_\mathrm{eff}$}
\newcommand{\feh}{\ensuremath{[\mbox{Fe}/\mbox{H}]}}
\newcommand{\logr}{log\big(\ensuremath{R'_{\mbox{\scriptsize HK}}}\big)}
\newcommand{\vsini}{$v \sin i$}
\newcommand{\Msini}{$M \sin i$}

\title{Friends of Hot Jupiters III: \\
   An Infrared Spectroscopic Search for Low-Mass Stellar Companions}

\author{Danielle Piskorz\altaffilmark{1}, Heather A. Knutson\altaffilmark{1}, Henry Ngo\altaffilmark{1}, Philip S. Muirhead\altaffilmark{2}, Konstantin Batygin\altaffilmark{1}, Justin R. Crepp\altaffilmark{3}, Sasha Hinkley\altaffilmark{4},  Timothy D. Morton\altaffilmark{5}}

\email{dpiskorz@gps.caltech.edu}
\altaffiltext{1}{Division of Geological and Planetary Sciences, California Institute of Technology, Pasadena, CA}
\altaffiltext{2}{Institute for Astrophysical Research, Boston University, Boston, MA}
\altaffiltext{3}{Department of Physics, University of Notre Dame, South Bend, IN}
\altaffiltext{4}{Department of Physics and Astronomy, University of Exeter, Exeter, UK}
\altaffiltext{5}{Department of Astrophysical Sciences, Princeton University, Princeton, NJ}

\begin{abstract}
Surveys of nearby field stars indicate that stellar binaries are common, yet little is known about the effects that these companions may have on planet formation and evolution. The Friends of Hot Jupiters project uses three complementary techniques to search for stellar companions to known planet-hosting stars: radial velocity monitoring, adaptive optics imaging, and near-infrared spectroscopy. In this paper, we examine high-resolution \textit{K} band infrared spectra of fifty stars hosting gas giant planets on short-period orbits. We use spectral fitting to search for blended lines due to the presence of cool stellar companions in the spectra of our target stars, where we are sensitive to companions with temperatures between 3500-5000 K and projected separations less than 100 AU in most systems. We identify eight systems with candidate low-mass companions, including one companion that was independently detected in our AO imaging survey. For systems with radial velocity accelerations, a spectroscopic non-detection rules out scenarios involving a stellar companion in a high inclination orbit. We use these data to place an upper limit on the stellar binary fraction at small projected separations, and show that the observed population of candidate companions is consistent with that of field stars and also with the population of wide-separation companions detected in our previous AO survey.  We find no evidence that spectroscopic stellar companions are preferentially located in systems with short-period gas giant planets on eccentric and/or misaligned orbits.
\end{abstract}

\section{Introduction}
Approximately 1$\%$ of nearby Sun-like stars host short-period gas giant planets, known as ``hot Jupiters" \citep{Wright2012}. Standard models of planet formation suggest that hot Jupiters are unlikely to have formed in situ, but must have formed beyond the ice line and migrated inward \citep{Pollack1996, Lin1996}. In this scenario, proposed migration models include both planet-disk (type II) interactions \citep{Goldreich980, Kley2012} and dynamical models including Kozai migration \citep{Malmberg2008, Fabrycky2007, Naoz2011}, planet-planet scattering \citep{Nagasawa2008, Beauge2012}, and secular chaos \citep{Wu2010}. While disk-driven migration is controlled primarily by local interactions, dynamical migration processes can be strongly affected by the presence of distant massive companions. In particular, the simplest variant of Kozai migration requires a perturbing star \citep{Wu2003}, while planet-planet scattering can in principle be triggered by external perturbations \citep{Batygin2011}. By studying the present-day properties of hot Jupiter systems, we can distinguish between competing formation and migration channels. 

We generally expect that in isolation disk migration should produce hot Jupiters on circular and well-aligned orbits, while dynamical migration simulations frequently result in planets with orbits that are eccentric and/or misaligned with respect to the star's spin axis.  Surveys of hot Jupiter spin-orbit alignments indicate that approximately half of all hot Jupiter systems are misaligned \citep{Winn2010,Triaud2010, Albrecht2012}, suggesting that three-body dynamics may play an important role in these systems. On the other hand, the apparent paucity of high eccentricity gas giant planets at intermediate orbital periods suggest that less than half of all hot Jupiters could have migrated via the star-planet Kozai-Lidov mechanism \citep{Dawson2013}. Alternatively, the presence of a stellar companion can also tilt the protoplanetary disk with respect to the stellar rotation axis, causing spin-orbit misalignments before planets have even formed \citep{Batygin2012, Spalding2014}. Regardless of whether it is the disk or the planet orbit being tilted, both scenarios require the presence of a massive outer companion on a non-coplanar orbit (albeit in different epochs) in order to explain the present-day spin-orbit misalignments observed in a significant fraction of hot Jupiter systems.

Although a majority of the extrasolar planets detected to date appear to orbit single stars, this is somewhat surprising as surveys of field stars indicate that approximately half of all Sun-like stars in the solar neighborhood are found in binaries \citep{Duquennoy1991, Raghavan2010}. It is unclear exactly what role a binary companion might play in the process of planet formation and migration. It has been suggested that wide separation binaries may warp or even truncate the outer edges of the protoplanetary disk and reduce average disk lifetimes (e.g., Terquem \& Bertout 1993; Pichardo et al. 2005; Kraus et al. 2012; Cheetham et al. 2015). Dynamical interactions with a distant companion may increase turbulent velocities in the protoplanetary disk, thereby preventing materials from condensing \citep{Mayer2005}. By searching for stellar companions to known planetary systems we can constrain their potential effects on these planetary systems, albeit with the caveat that close encounters between stars forming in crowded cluster environments may have similar effects (e.g., Bonnell et al. 2001; Spurzem et al. 2009; Hao et al. 2013; Zheng et al. 2015).

Previous surveys have identified a number of stellar companions in known planetary systems \citep{Eggenberger2007, Raghavan2010, Wang2014a}, but only a handful are close binaries with hot Jupiters orbiting the primary star. The \textit{Kepler} mission has detected approximately a dozen circumbinary planets to date (e.g., \citealt{Doyle2011, Welsh2012}) and a number of adaptive optics (AO) surveys have proven to be effective at detecting more widely separated stellar companions (\citealt{Wang2015, Woellert2015a, Woellert2015}; see \citealt{Ngo2015} for a complete review of surveys prior to 2015). The Friends of Hot Jupiters (FOHJ) project systematically tests the validity of dynamical models of hot Jupiter migration and performs a dedicated inquest on the stellar multiplicity rate of hot Jupiter systems. We focus on a sample of nearby transiting hot Jupiters with well-characterized spin-orbit alignments and orbital eccentricities, divided into a control group with circular, well-aligned orbits and an experimental group with eccentric and/or misaligned orbits. Our approach differs from that of most previous surveys, which typically focused on either non-transiting planets or transiting planet candidates in the \textit{Kepler} sample, of which the vast majority are too small or too faint to detect the Rossiter-McLaughlin effect and measure their corresponding spin-orbit misalignments (e.g., Lillo-Box 2012; Adams et al. 2012). 

In \cite{Knutson2014} we searched for long-term radial velocity accelerations due to distant planetary or stellar companions in these systems, and found that  51$\pm$10$\%$ of the stars in our sample hosted planetary mass companions with orbits between 1-20 AU. In \cite{Bechter2014} and \cite{Ngo2015} we performed a complementary \textit{K} band AO imaging search for stellar companions on relatively wide orbits, and found a binary rate of 48$\pm$9$\%$  for stellar companions with projected separations between 50-2000 AU. This rate is approximately twice that of field stars having companions in this semi-major axis range \citep{Raghavan2010}, suggesting that stellar companions may play a role in the formation of these systems. Although previous imaging studies hinted at a high stellar multiplicity rate for transiting planet host stars (see Ngo et al. 2015 for a complete review), our study was the first to confirm that the imaged companions were gravitationally bound and to derive a completeness-corrected multiplicity rate for hot Jupiter host stars. In both surveys there was no indication that eccentric or misaligned systems were more likely to have a massive outer companion than their circular and well-aligned counterparts.

In this study, we use Keck NIRSPEC (Near InfraRed Echelle SPECtrograph; McLean et al. 1998) to search for stellar companions that might have gone undetected in our AO and radial velocity observations. We use high-resolution \textit{K} band spectroscopy to search for blended lines from cool stellar companions, exploiting the deep CO molecular absorption features present in cool stars and distinct from the lines of the hotter primaries. We expect that these companions will have relatively small projected separations and/or high orbital inclinations, in order to be consistent with our previous radial velocity and AO observations of these systems. For systems in which we detect companions, we can estimate their effective temperatures and place an upper limit on their projected separations from the primary.  

A number of previous studies have used high-resolution spectroscopy to locate hidden binary companions. \cite{Gagliuffi2014} analyzed a sample of 815 M and L dwarf spectra taken with IRTF SpeX in order to locate blended stellar companions with relatively low effective temperatures (also see Burgasser et al. 2010a). \cite{Guenther2013} used VLT CRIRES to identify approximately twenty planet-hosting stars in the \textit{CoRoT }sample with blended spectra from close stellar companions. \cite{Kolbl2015} observed planet-hosting stars from the \textit{Kepler} survey with optical Keck HIRES spectroscopy in order to search for binary companions having relative radial velocities greater than 10 km/s such that the secondary absorption lines are Doppler shifted. A similar technique is used to probe absorption lines in the atmospheres of hot Jupiters, which exhibit rapidly varying velocity offsets \citep{Snellen2010, deMooij2012, Birkby2013, Lockwood2014}.  For cases where the hidden companion has a significantly different effective temperature than the target star, \cite{Kolbl2015} were also able to detect companions with smaller radial velocity offsets. Of the 1160 \textit{Kepler} stars with candidate transiting planets, sixty-three showed spectroscopic evidence for a companion star.  We use a similar approach in our survey, but observe in the infrared in order to increase our sensitivity to relatively cool stellar companions.

In Section 2 we present a description of our observations and subsequent model fitting, and in Section 3 we discuss the resulting spectroscopic detections. In Section 4 we compare our results to those of the adaptive optics and radial velocity portions of the Friends of Hot Jupiters survey. In Section 5 we compute the companion fraction for our sample.

\section{Observations and Data Analysis}
\begin{figure}[t]
\centering
\noindent\includegraphics[width=20pc]{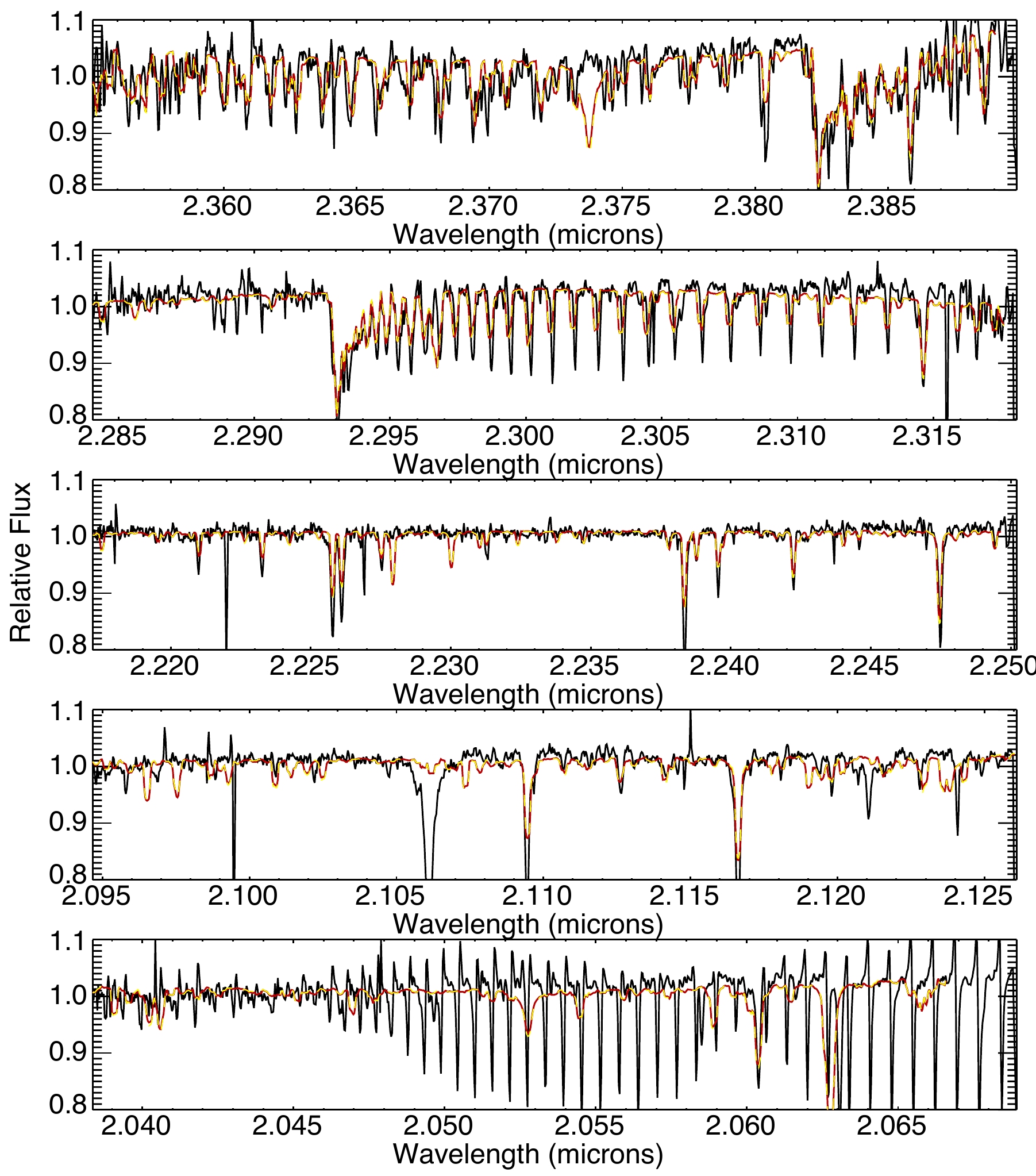}
\caption{Telluric-corrected and wavelength-calibrated \textit{K} band spectrum for WASP-2 taken with NIRSPEC on UT July 4 2013 shown in black with the best-fit one-star model overplotted with a yellow dashed line and the best-fit two-star model overplotted with a red solid line. ``Chirps" in the data, especially from 2.040-2.065 $\mu$m are due to incomplete telluric removal.}
\label{spectrum}
\end{figure}

We observed fifty short-period transiting gas giant planetary systems on four separate nights (UT August 27 2012, January 28 2013, March 2 2013, and July 4 2013) using NIRSPEC at the W.M. Keck Observatory on Mauna Kea, which has a resolution $R$ = 30,000 in the \textit{K} band (2.0-2.4 microns). See \cite{Knutson2014} for details on the sample selection. We extract one-dimensional spectra from the raw images using an IDL (Interactive Data Language) pipeline that flat fields and dark subtracts the images as well as removes any bad pixels following the methods described in \cite{Boogert2002}. We correct for telluric absorption by dividing the science target spectrum by that of a calibrator star with an intrinsically flat spectrum, usually a nearby rapidly-rotating A-star, at a similar air mass on the same night, where we have empirically shifted the calibrator spectrum to match the wavelength solution of each target star.  As an example, the telluric-corrected and wavelength-calibrated spectrum for WASP-2 taken on UT July 4 2013 is shown in Figure~\ref{spectrum}. 

\subsection{One-Star Model}
\label{onestar}
We fit each \textit{K} band spectrum with a PHOENIX stellar model \citep{Husser2013} interpolated to match the published effective temperature $T_\mathrm{eff}$, surface gravity $\mathrm{log}(g)$, and metallicity \feh~of the target star. See Table \ref{targetlist} for a list of targets and their stellar properties. In accordance with \cite{Gray2005}, the synthetic spectrum is rotationally broadened by convolving it with the following kernel:
\begin{equation}
\footnotesize
G(\Delta\lambda)=\frac{2(1-\epsilon)\left(1-\Delta\lambda^2\right)^{1/2}+\frac{1}{2}\pi\epsilon\left(1-\Delta\lambda^2\right)}{\pi c \left(1-\frac{\epsilon}{3}\right)}
\end{equation}
where $\Delta \lambda = \frac{\lambda v\mathrm{sin}i}{c}$, $v \sin i$ is the line-of-sight rotational velocity as listed in Table ~\ref{targetlist}, and $\epsilon$ is the limb darkening coefficient of the target taken from \cite{Claret2000}. 

\begin{deluxetable*}{lccccccccc}
\centering
\tabletypesize{\scriptsize}
\tablewidth{0pt}
\tablecaption{Target List and Stellar Properties}
\tablehead{ \colhead{Target Star} &\colhead{$R$ (R$_{\bigodot}$) } &\colhead{\teff (K)} & \colhead{$D$ (pc)} & \colhead{log$g$} &\colhead{\feh} &\colhead{\vsini (km/s)} &\colhead{$\epsilon$} &\colhead{\logr} & \colhead{References}}
\startdata

GJ 436     & 0.45 & 3416 		& 10.14 & 4.83 & -0.03 & 3.0 & 0.3063 & -5.298 		&  1, 2, 3, 4\\
HAT-P-2  & 1.64 & 6411 		& 125.3 & 4.16 & 0.14 & 20.8 & 0.2828 & -4.7 		&   5, 6\\
HAT-P-4  & 1.46 & 6687		& 293.5 & 4.14 & 0.2 & 5.6 & 0.2697 & -5.082 		&  5, 7, 8\\
HAT-P-6  & 1.46 & 6687 		& 277.8 & 4.22 & -0.11 & 8.9 & 0.2556 & -4.799 	& 5, 9\\
HAT-P-7  & 1.90 & 6259 		& 320 & 4.02 & 0.15 & 5 & 0.2900 & -5.018		 & 5, 10, 11\\
HAT-P-8  & 1.48 & 6223 		& 230 & 4.19 & -0.04 & 12.6 & 0.2917 & -4.985 	&  5, 12, 13\\
HAT-P-10  & 0.79 & 4974 		& 121.7 & 4.56 & 0.25 & 1.9 & 0.3613 & -4.823 		&  5, 14\\
HAT-P-11  & 0.75 & 4792 		& 38.0 & 4.59 & 0.33 & 0.8 & 0.3789 & -4.567 		&  5,15\\
HAT-P-12  & 0.70 & 4650 	& 139.1 & 4.61 & -0.29 & 0.5 & 0.3932 & -5.104 		& 16\\
HAT-P-13  & 1.76 & 5720 	& 214 & 4.13 & 0.46 & 3.1 & 0.3117 & -5.138 		&  5, 14, 17\\
HAT-P-14  & 1.59 & 6671 	& 205 & 4.25 & 0.07 & 9.0 & 0.2714 & -4.855 		&  5, 17, 18\\
HAT-P-15  & 1.08 & 5640 		& 190 & 4.38 & 0.31 & 2.1 & 0.3229 & -4.977 		&   5, 19\\
HAT-P-16  & 1.24 & 6140 	& 235 & 4.34 & 0.12 & 3.4 & 0.2958 & -4.863 	&   5, 20\\
HAT-P-17  & 0.84 & 5345 	& 90 & 4.53 & 0.06 & 1.3 & 0.3349 & -5.043 		&  5, 21\\
HAT-P-18  & 0.75 & 4790 	& 166 & 4.56 & 0.14 & 0.8 & 0.3789 & -4.799 		&  5, 22\\
HAT-P-20  & 0.70 & 4619 	& 70 & 4.64 & 0.26 & 2.6 & 0.3932 & -4.506 		&  5, 23\\
HAT-P-22  & 1.04 & 5367 		& 82 & 4.37 & 0.29 & 1.5 & 0.3349 & -4.901 		&  5, 23\\
HAT-P-24  & 1.32 & 6329 	& 396 & 4.27 & -0.21 & 11.4 & 0.2871 & -4.955 	&  5, 24\\
HAT-P-26  & 0.79 & 5142 	& 134 & 4.56 & 0.1 & 1.4 & 0.3478 & -5.008 		&  5, 22\\
HAT-P-29  & 1.22 & 6086 	& 322 & 4.34 & 0.14 & 4.4 & 0.2982 & -5.096 	& 5, 25\\
HAT-P-30  & 1.22 & 6304 	& 193 & 4.36 & 0.13 & 2.2 & 0.2882 & 5.169 	& 26\\
HAT-P-31  & 1.36 & 6065 		& 354 & 4.26 & 0.15 & 0.5 & 0.2992 & 5.169 	& 27\\
HAT-P-32  & 1.22 & 6207 	& 283 & 4.33 & -0.04 & 20.7 & 0.2927 & -4.641 	& 22\\
HAT-P-33  & 1.64 & 6446	 	& 387 & 4.15 & 0.07 & 13.7 & 0.2812 & -4.87 	& 22\\
HAT-P-34  & 1.56 & 6442 	& 257 & 3.98 & 0.21 & 24.5 & 0.2814 & -4.931 	&  28\\
HD 149026  & 1.53 & 6103 	& 80.8 & 4.27 & 0.24 & 6.3 & 0.2975 & -5.03 	& 5, 29\\
TrES-2  & 0.95 & 5850 		& 220 & 4.47 & -0.01 & 0.8 & 0.3114 & -4.949 		&  5, 30, 31\\ 
TrES-3   & 0.83 & 5514 		& 258.5 & 4.57 & -0.2 & 1.3 & 0.3229 & -4.549 		& 5, 32\\
TrES-4  & 1.83 & 6200 		& 476 & 4.05 & 0.14 & 8.5 & 0.2928 & -5.104 		& 32\\
WASP-1  & 1.50 & 6160 		& 380 & 4.21 & 0.14 & 1.7 & 0.2947 & -5.114 		& 5, 17\\
WASP-2  & 1.06 & 5255 		& 140 & 4.52 & 0.06 & 1.9 & 0.3478 & -5.054 		&  5, 17, 33\\
WASP-3  & 1.21 & 6375 		& 220 & 4.28 & -0.06 & 15.4 & 0.2850 & -4.872 	&  5, 34, 35\\
WASP-4  & 0.90 & 5540 		& 280.9 & 4.47 & 0 & 3.4 & 0.3229 & -4.85		 & 5, 36, 37\\
WASP-7   & 1.32 & 6520		& 140 & 4.32 & 0 & 18.1 & 0.2783 & -4.8 		&  37, 38\\
WASP-8  & 1.05 & 5570 		& 87 & 4.40 & 0.17 & 2.7 & 0.3114 & -4.709 		&  37, 39\\
WASP-10  & 0.70 & 4735 	& 90 & 4.51 & 0.05 & 2.9 & 0.3789 & -4.704		&  5, 40, 41\\
WASP-14  & 1.67 & 6462 	& 160 & 4.29 & -0.13 & 3.5 & 0.2810 & -4.923 	& 5, 17, 42\\
WASP-15  & 1.52 & 6405 	& 256 & 4.40& 0 & 4.9 & 0.2836 & -5.286 		& 37, 43\\
WASP-16  & 1.09 & 5630 	& 174 & 4.21 & 0.07 & 2.5 & 0.3233 & -5.048 		& 37, 43\\
WASP-17  & 1.58 & 6550	 	& 476 & 4.14 & -0.02 & 9.8 & 0.2763 & -5.331		& 5, 17, 37, 44\\
WASP-18  & 1.29 & 6368 		& 122.6 & 4.37 & 0.11 & 10.9 & 0.2853 & -5.43		&  5, 37, 38\\
WASP-19  & 1.02 & 5460 	& 250 & 4.50 & 0.05 & 4.5 & 0.3349 & -4.66 			& 45 \\
WASP-22  & 1.22 & 5958 	& 300 & 4.50 & 0.05 & 4.5 & 0.3041 & -5.065 		& 45 \\
WASP-24  & 1.33 & 6107 	& 332.5 & 4.26 & -0.02 & 6.1 & 0.2973 & -5.139 	&  5, 46\\
WASP-34  & 0.93 & 5700 		& 120 & 4.50 & -0.02 & 1.4 & 0.3114 & -5.163 		&  47 \\
WASP-38  & 1.35 & 6187 		& 110 & 4.25 & -0.02 & 8.6 & 0.2936 & -5.158 	&  5, 48\\
XO-2  & 0.97 & 5377 		& 156.0 & 4.45 & 0.35 & 1.0 & 0.3349 & -4.988		& 5, 49\\
XO-3  & 1.38 & 6759 		& 185.7 & 4.24 & -0.05 & 20.3 & 0.2664 & -4.595	& 5, 50, 51\\
XO-4  & 1.56 & 6297 		& 308.2 & 4.17 & -0.03 & 8.8 & 0.2882 & -5.292	& 5, 52\\
XO-5  & 1.08 & 5370 		& 260 & 4.31 & 0.05 & 0.7 & 0.3349 & -5.147 		& 11\\

\enddata
\tablecomments{Distances estimated from stellar models. All $\epsilon$ values from \cite{Claret2000}. All log$R_{HK}$ values from \cite{Knutson2010}. WASP-12 was in the original Friends of Hot Jupiter sample, but eliminated from the NIRSPEC survey because of its low elevation at the time of observation.}
\tablerefs{(1) \cite{Bonfils2005}; (2) \cite{Maness2007}; (3) \cite{VonBraun2012}; (4) \cite{Torres2007}; (5) \cite{Torres2012}; (6) \cite{Pal2010}; (7) \cite{Winn2011}; (8) \cite{Kovacs2007}; (9) \cite{Noyes2008}; (10) \cite{Eylen2012}; (11) \cite{Pal2009}; (12) \cite{Mancini2013}; (13) \cite{Latham2009}; (14) \cite{Bakos2009}; (15) \cite{Bakos2010}; (16) \cite{Hartman2009}; (17) \cite{Southworth2012}; (18) \cite{Torres2010}; (19) \cite{Kovacs2010}; (20) \cite{Buchhave2010}; (21) \cite{Howard2012}; (22) \cite{Hartman2011}; (23) \cite{Bakos2011}; (24) \cite{Kipping2010}; (25) \cite{Buchhave2011}; (26) \cite{Johnson2011}; (27)  \cite{Kipping2011}; (28) \cite{Bakos2012};  (29) \cite{Carter2009}; (30) \cite{Barclay2012}; (31) \cite{O'Donovan2006}; (32) \cite{Sozzetti2009}; (33) \cite{Bergfors2012}; (34) \cite{Miller2010}; (35) \cite{Gibson2008}; (36) \cite{Wilson2008}; (37) \cite{Doyle2012}; (38) \cite{Hellier2009}; (39) \cite{Queloz2010}; (40) \cite{Johnson2009}; (41) \cite{Christian2009}; (42) \cite{Joshi2009}; (43) \cite{Southworth2013}; (44) \cite{Anderson2010}; (45)  \cite{Anderson2011}; (46) \cite{Street2010}; (47) \cite{Smalley2011}; (48) \cite{Brown2012};  (49) \cite{Burke2007}; (50) \cite{Winn2008}; (51) \cite{Johns-Krull2008}; (52) \cite{McCullough2008}}
\label{targetlist}
\end{deluxetable*}

\subsection{Two-Star Model}
We construct a two-star model by combining the rotationally-broadened PHOENIX model appropriate for the target star with another PHOENIX model corresponding to a faint cool companion in the system. For each target, we create 34 two-star models, each with a different companion effective temperature ranging from 2300 to 5500 K. We assume all our companion stars have $\mathrm{log}(g)=5.00$.  We also assume our companion stars have the same radial veolcities as the primary stars because companions in short-period orbits would already have been detected in our radial velocity observations of the primary star. For example, a K-dwarf companion to a typical star in our survey with a random orbital orientation must on average be located beyond 10 AU in order to avoid creating a detectable RV signal.  At this separation, the companion star would have a RV offset of 6 km/s, corresponding to 0.4 pixels in our NIRSPEC observations. This choice represents a departure from traditional spectroscopic binary analyses (e.g., Zucker \& Mazeh 1994), which allow for an arbitrary radial velocity offset between the two binary components.  Although our decision to fix the radial velocity offset between the two stars to zero precludes us from detecting chance blends with unassociated background or foreground stars, we note that such blends would need to have a differential magnitude less than 5.0 in order to be detectable and a separation of less than 0.4" in order to fall within our slit.  In our AO survey of these stars we found that all candidate stellar companions with a differential K band magnitude less than 6.0 located within 5" of the primary were in fact bound companions (see Fig. 4 in Ngo et al. 2015), and we therefore consider it unlikely that any chance blends would occur in our sample that meet the above criteria.

We set \feh~= 0 for our companion stars as the primary stars in these systems all have near-solar or solar metallicities. We evaluate the effect of metallicity on our models by re-running our fits to the most metal-rich star in our sample (HAT-P-13, \feh~= 0.46 $\pm$ 0.07), and find that our results are indistinguishable from those of stellar metallicity models. Since cool stars typically have $v \sin i$ values less than 5 km/s, instrumental broadening will dominate and we fix the rotational broadening to zero for our cool star companion models. PHOENIX models are given in units of flux per unit surface area, and we multiply the spectra of the primary and companion stars by their respective areas in order to convert to total flux. We take the value for the radius of the primary star from the published literature, and we calculate the radius of the companion as a function of its effective temperature using the stellar evolution models of \cite{Baraffe2003}.

\subsection{Fitting Procedure and Detection Metric}
\label{fit}
We first fit the one-star model to the calibrated data, assuming constant errors at each wavelength bin. The wavelength solution (described to third order as $\lambda=ax^2+bx+c$ where $x$ is pixel number and $a$, $b$, and $c$ are free parameters) and the width of the instrumental broadening kernel are left as free parameters when fitting each individual spectrum. We allow the instrumental broadening to vary across all orders and find that it remains roughly consistent throughout (full-width half-maximum $\sim$ 0.05 cm$^{-1}$). The instrumental broadening kernel is assumed to be Gaussian and represents the effect of poor seeing and the interaction of the starlight with the instrumental apparatus. The instrumental broadening varies from target to target according to the orientation of the telescope and the air mass of the observations. We find the best-fit model by minimizing $\chi^{2}$ and check that we have found the correct global minimum by repeating the calculation with different initial guesses in the parameter space.

We then use the best-fit one-star model to determine empirical error bars for the data. These error bars are calculated as the standard deviation of each residual and its twenty nearest neighbors. This method of error calculation allows us to directly estimate the combined error due to the calibration, model fit, and photon noise contributions, many of which are difficult to predict a priori. We note that these empirically determined error values are only an approximation to the true error distribution; our use of the $\chi^{2}$ metric implicitly assumes that each wavelength measurement is drawn from an independent Gaussian distribution with a width $\sigma$ determined by our empirical estimates. This is fundamentally an approximation and we therefore use the $\chi^{2}$ values from our fits as a metric of relative goodness-of-fit rather than an absolute measurement of the probability of a given model. With these new error bars, we refit the best-fit one-star model at the best-fitting grid point. We use the final best-fit parameters for the one-star model as the initial guess for fitting the two-star model. Although we allow the instrumental broadening and wavelength solution to vary between the one and two-star models, we find consistent values between the two versions of the fit. We exclude the third order (2.155-2.185 $\mu$m) from our analysis, where we find that our telluric A star standards have a strong Brackett gamma absorption line that propagates into our target spectra when applying our telluric correction. 

\begin{figure*}[t]
\centering
\noindent\includegraphics[width=45pc]{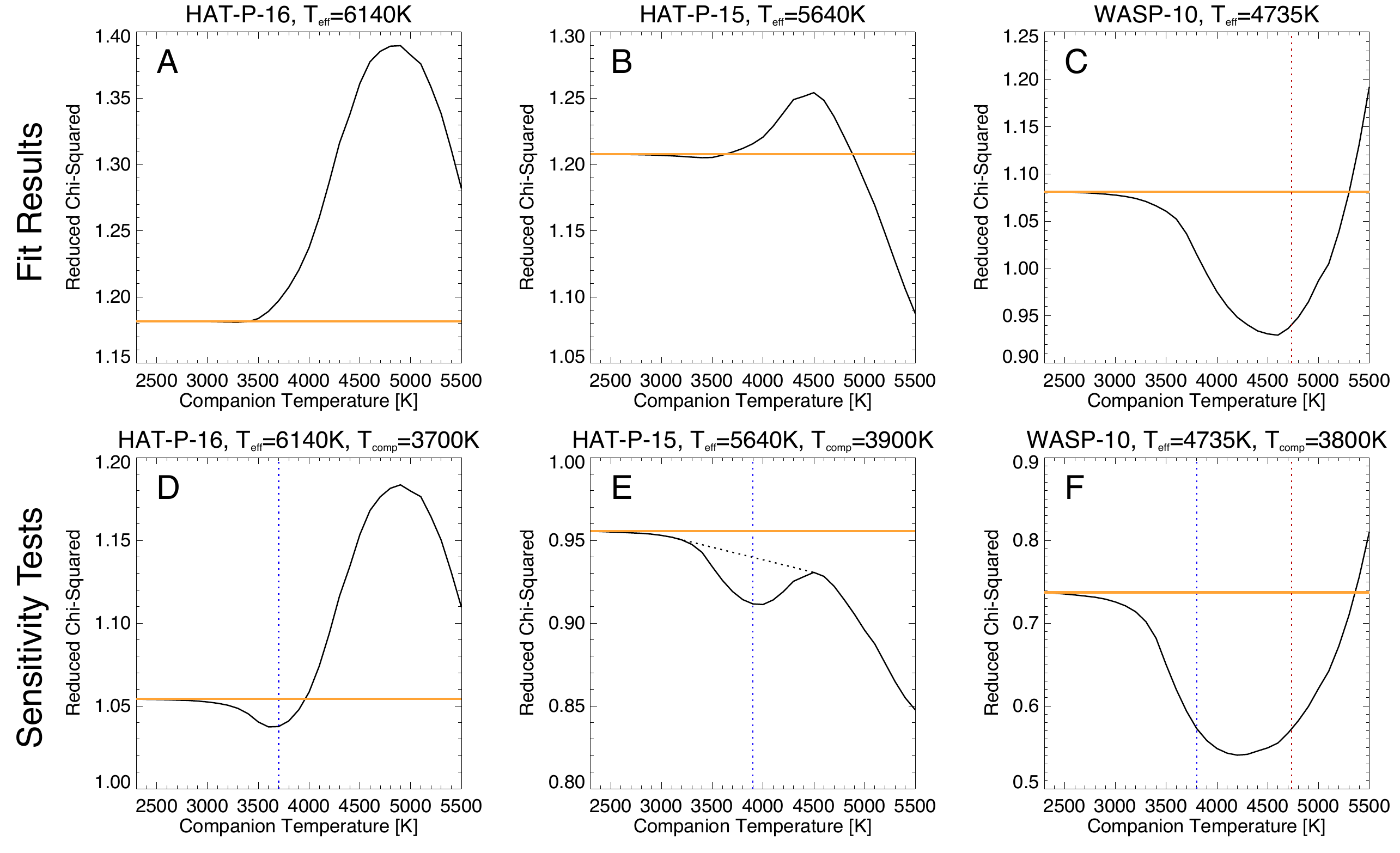}
\caption{ Fit results (panels A - C) and sensitivity tests (panels D - F) for HAT-P-16, HAT-P-15, and WASP-10. The solid orange and black lines represent the reduced chi-squared value of the one- and two-star models, respectively. The red dotted line represents the effective temperature of the host star and the blue dotted lines in panels D-F represent the effective temperature of the injected stellar companion. The dotted black line in panel E represents the expected slope of the $\chi^{2}$ trend for a two-star fit in the case where no companion was present. These systems are classified as non-detections since there is no reliable reduction in $\chi^{2}$ when a second star having an effective temperature distinct from that of the target star is added to the model fit. We are able to inject and succesfully recover signals due to 3700 K and 3900 K companions in the HAT-P-16 and HAT-P-15 systems (panels D and E, respectively). We cannot recover the 3800 K signal injected into the WASP-10 system due to WASP-10's low effective temperature. }
\label{nondetection}
\end{figure*}

We plot the reduced chi-squared ($\chi^{2}_{red}$) value for the two-star fit as a function of the stellar companion temperature and look for minima indicating the presence of a cool companion.  The $\chi^{2}_{red}$ for the coolest stellar companions approaches that of the single-star fit, indicating that we are not sensitive to companions below a certain temperature, as shown in panels A-C of Figure~\ref{nondetection}. 

We also compare the one- and two-star fits using the Bayesian Information Criterion (BIC), which is defined as: 
\begin{equation}
\mathrm{BIC} = \chi^{2}+N\mathrm{ln}(n)
\label{BIC}
\end{equation}
where $\chi^{2}$ is the canonical chi-squared value, $N$ is the number of free parameters, and $n$ is the number of data points.  For our purposes, N=4 for the one-star model, N=5 for the two-star model, and n = 4980. In order to be classified as a detection, there must be a significant improvement in the $\chi^{2}_{red}$ and in the BIC for the two-star model, and we must be able to verify the presence of absorption lines from the cool companion that are distinct from those of the primary (i.e., the code is not just improving the fit to a single star spectrum by overlaying a second nearly identical spectrum and better fitting to the measured line profiles). We find that in all cases the BIC gives results that are equivalent to the $\chi^{2}_{red}$ approach. For the systems where we detect candidate stellar companions, we list the effective temperature of the cool companion that produces the largest improvement in $\chi^{2}_{red}$ over the one-star model. In some cases there is a broad minimum in $\chi^{2}_{red}$ for the two-star model centered on the effective temperature of the primary, which can create a slope that extends out to relatively low companion temperatures. We correct for the effect of this slope in cases where we detect a candidate stellar companion at lower temperatures by interpolating the slope across the region spanned by the minimum due to the companion and subtracting the interpolated trend.  The endpoints for this interpolated line are chosen by finding the locations on either side of the local reduced chi-squared minimum having slopes equal to within 10\%.

Our wavelength-dependent measurement errors are determined empirically from the data themselves and we expect them to be dominated by systematic errors due to imperfect corrections for telluric absorption and in our stellar models. As a result, there is no formal metric for determining whether or not a detection is statistically significant, and we therefore rank-order our detections from strongest to weakest according to the depth of the minimum in $\chi^{2}_{red}$.  Fitting our data with this same procedure, but with uniform error bars, gives the same final sets of detections and non-detections, with only slight variations in detection strengths and companion effective temperatures.

We estimate the uncertainties on the effective temperatures of the candidate stellar companions using two methods. We first calculated the range in effective temperature corresponding to a 1$\sigma$ change in our best-fit $\chi^{2}$ value, but found that this method produced unrealistically small error bars with a typical size of 120 K. We instead adopted a more conservative method in which we calculate the range in effective temperature corresponding to a change in $\chi^{2}_{red}$ equal to half the total difference between the one- and two-star models at the best-fit companion temperature. We find typical uncertainties of 250 K using this method, as shown in Table~\ref{detection}. Figure~\ref{spectrum} shows the best-fit one-star and two-star models for WASP-2 in yellow and red, respectively.

\subsection{Sensitivity Tests}
\label{sensitivity}
We evaluate our sensitivity to cool stellar companions in individual systems with non-detections by injecting synthetic companions into our data and determining the lowest effective temperature for which we can reliably detect the injected companion.  In doing so, we characterize our dispositive null detections, where our lack of detection implies that there is no companion in a specific temperature and semi-major axis range in that system.  We create each synthetic companion spectrum by applying the previously calculated best-fit wavelength solution and instrumental broadening to a PHOENIX spectrum for a stellar companion at the desired temperature.  We add this fake spectrum to the target data, scaled to the band-integrated flux of the primary star according to 
\begin{equation}
F= \frac{R_{p}^{2}\int_{\lambda_1}^{\lambda_2} \! I(T_p, \lambda)  \,  \mathrm{d}\lambda}{R_{c}^{2}\int_{\lambda_1}^{\lambda_2} \! I(T_c, \lambda)  \,  \mathrm{d}\lambda}
\label{fluxfactor}
\end{equation}
where  $R_p$ is the radius of the primary star, $R_s$ is the radius of the companion star, $\lambda_1$ is the short-wavelength limit of the \textit{K} band, $\lambda_2$ is the long-wavelength limit of the \textit{K} band, $I(T, \lambda)$ is the surface brightness of the PHOENIX spectrum for each target star and each synthetic companion, $T_p$ is the effective temperature of the primary star, and $T_c$ is the effective temperature of the companion star. We run this composite spectrum through the fitting procedure described in Sections~\ref{onestar} -~\ref{fit} and calculate a corresponding lower limit on the temperature of the stellar companions that can be detected in our data. Note that the properties of the hot Jupiter host stars are known in advance from high-resolution optical spectroscopy, which will be minimally affected by contamination from an M dwarf companion.  This precludes a scenario in which a G+M star spectrum is mistaken for a K star spectrum, as degeneracies are only possible when the temperatures of both the primary and companion star are allowed to very as free parameters in the fits.

We carry out the procedure described above on the targets having \teff~$\le$ 5700 K and report the dispositive null detections in Table~\ref{non}. For targets having \teff~$>$ 5700 K (which all have chi-squared curves shaped similarly to Panel A of Figure~\ref{nondetection}), we find that the range of effective temperatures where the difference in $\chi^{2}_{red}$ between the one- and two- star models is greater than 0.005 is the same as the range of effective temperatures suggested by the full injection-and-recovery method. For these targets, we use this threshold in $\Delta \chi^{2}_{red}$ rather than running the full sensitivity test on each individual system, and report the range of companion temperatures corresponding to dispositive null detections in Table~\ref{non}.

\section{Results}
The results of our analysis are shown in Tables~\ref{non} and~\ref{detection}. Table~\ref{non} lists systems with non-detections as well as the range of companion temperatures that can be ruled out for systems with non-detections. Table~\ref{detection} is organized according to the strength of the detection, which we define as the improvement in $\chi^{2}$ for the two-star model as compared to the one-star model. 

\begin{deluxetable}{lcc} 
\tablewidth{0pt}
\tablecaption{NIRSPEC Sensitivity Limits for Systems with  Dispositive Null Detections}

\startdata
\hline \hline
\vspace{0.01cm}
Target Star & $T_{\mathrm{eff}}$ Range & Max. Sep.)\tablenotemark{a} \\
 & (K) & (AU) \\
\hline
\vspace{0.02cm}

HAT-P-2&	3900--5500& 49	\\
HAT-P-6&	3400--5500&	112	\\
HAT-P-7&	3500--5500&	138\\
HAT-P-8&	3600--5500&	99	\\
HAT-P-14 &	3700--5500&	89	\\
HAT-P-15\tablenotemark{b}&	3400--4500&	82	\\
HAT-P-16	&3600--5500&	101	\\
HAT-P-24&	3600--5500&171		\\
HAT-P-29&	3600--5500&	139	\\
HAT-P-30&	3800--5500&	83	\\
HAT-P-31&	3900--5500&	153	\\
HAT-P-32&	3500--5500&	122	\\
HAT-P-33&	3800--5500&	167	\\
HD 149026&	3800--5500&	34	\\
TrES-2&	3600--5500&	99	\\
TrES-3\tablenotemark{b}&	3600--5500&	98	\\
TrES-4&	4000--5500&	213	\\
WASP-1&	3800--5500&	164	\\
WASP-3	&3800--5500& 95		\\
WASP-4\tablenotemark{b}	&3700--4500& 130		\\
WASP-7&	3400--5500&	61	\\
WASP-14&	3500--5500&	69	\\
WASP-15&	3500--5500&	110	\\
WASP-16\tablenotemark{b}&	3300--4000&	75	\\
WASP-17&	3600--5500&	172	\\
WASP-18&	3900--5500&	43	\\
WASP-24&	3700--5500&	128	\\
WASP-34\tablenotemark{b}&	3300--3900&	52	\\
WASP-38&	3800--5500&	48	\\
XO-3&	3600--5500&	80	\\
XO-4&	3600--5500&	127	\\
\enddata
\label{non}
\tablenotetext{a}{This is the approximate maximum separation probed based on the size of the NIRSPEC slit (0.4'') and the system's parallax as given in Table~\ref{targetlist}.} 
\tablenotetext{b}{These targets have effective temperatures between 5500-5700 K. We are only sensitive to companions with effective temperatures at least 500 K cooler than the primary.}
\end{deluxetable}

\begin{deluxetable}{lcccc} 
\tablewidth{0pt}
\tablecaption{Systems with Candidate Companion Detections}
\startdata
\hline \hline 
\vspace{0.01cm}
Target Star & $\Delta\chi^{2}_{red}$ &\textbf{ $\Delta$BIC} & $T_{\mathrm{comp}}$ & Max. Sep.\tablenotemark{a} \\
& & & (K) & (AU) \\
\hline
\sidehead{\textbf{Detections ($\Delta\chi^{2}_{red} \ge $0.005\tablenotemark{b})}}
HAT-P-17  &  0.0162  & 72 & 3900$^{+200}_{-300}$   & 36 \\
WASP-2\tablenotemark{c}    &   0.0109 & 46  & 3800$^{+300}_{-350}$   & 56 \\
HAT-P-22  &   0.0105  & 44& 4000$^{+250}_{-400}$   & 33 \\
HAT-P-10\tablenotemark{d}   &  0.0099 &  41 & 4000$^{+200}_{-200}$  & 49 \\
HAT-P-26   &  0.0091 & 37  & 4000$^{+100}_{-350}$  & 54 \\
HAT-P-18\tablenotemark{d}   &  0.0085 & 34  & 4000$^{+200}_{-200}$  & 66 \\
HAT-P-13   &  0.0073 &28  & 3900$^{+300}_{-350}$  & 86 \\
HAT-P-34   &   0.0073 &28  & 3600$^{+150}_{-250}$  & 103 \\
WASP-22   &   0.0063  &23 & 3700$^{+150}_{-300}$  & 120 \\
XO-5           &    0.0050 & 16 & 3500$^{+250}_{-150}$ & 104\\
\sidehead{\textbf{Detections below empirical threshold for significance}}
\sidehead{(0.005 \textgreater  $~\Delta\chi^{2}_{red}$ \textgreater~0.0036\tablenotemark{e})}
HAT-P-4   &   0.0043 &13  & 3900$^{+450}_{-400}$ & 125\\
WASP-8   &  0.0043 & 13 & 3600$^{+350}_{-250}$  & 35 \\
\enddata
\label{detection}
\tablenotetext{a}{This is the approximate maximum projected separation based on the size of the NIRSPEC slit and the system's parallax as given in Table~\ref{targetlist}.} 
\tablenotetext{b}{This lower limit was chosen using WASP-2 as a benchmark becasue we independently detect the companion in our AO images. (See Section~\ref{WASP-2}.)}
\tablenotetext{c}{This detection is verified by our AO survey \citep{Ngo2015}, as discussed in Section~\ref{WASP-2}.}
\tablenotetext{d}{These detections are likely due to star spots rather than the presence of a companion, as discussed in Sections~\ref{HAT-18} and~/ref{HAT-10}.}
\tablenotetext{e}{This lower limit was chosen as there is a relatively large gap (0.001 in $\Delta\chi^{2}_{red}$) in $ \Delta\chi^{2}_{red}$ between this detection and the next non-detection.}
\end{deluxetable}

We classify systems with negligible improvements in $\chi^{2}$  for the two-star fit as non-detections. An example is shown in panel A of Figure~\ref{nondetection}. As discussed in Section~\ref{fit}, for targets having effective temperatures less than approximately 5500 K our two-star fit always finds a minimum corresponding to a fit where the primary and secondary stars have the same effective temperature. In practice this means that we are only sensitive to companions with effective temperatures at least 500 K cooler than that of the primary as illustrated in panels B and C of Figure~\ref{nondetection}.

Also shown in Figure ~\ref{nondetection} are the results of our sensitivity tests. Panels D and E shows the injection and recovery of 3700 K and 3900 K spectra in the HAT-P-16 and HAT-P-15 systems, respectively. There are 31 stars in our sample with similar sensitivity, for which we are sensitive to companions over the range of temperatures indicated in Table~\ref{non}. Seven of the coolest stars in our sample (WASP-10, GJ436, HAT-P-11, HAT-P-12, HAT-P-20, WASP-19, and XO-2) have effective temperatures between 3400-5500 K. For these seven targets, we are unable to detect injected companions at any of the temperatures considered in this study. Panel F of Figure~\ref{nondetection} shows our inability to recover a 3800 K spectrum injected into the WASP-10 system.

\begin{figure*}[b]
\centering
\noindent \includegraphics[width=42pc]{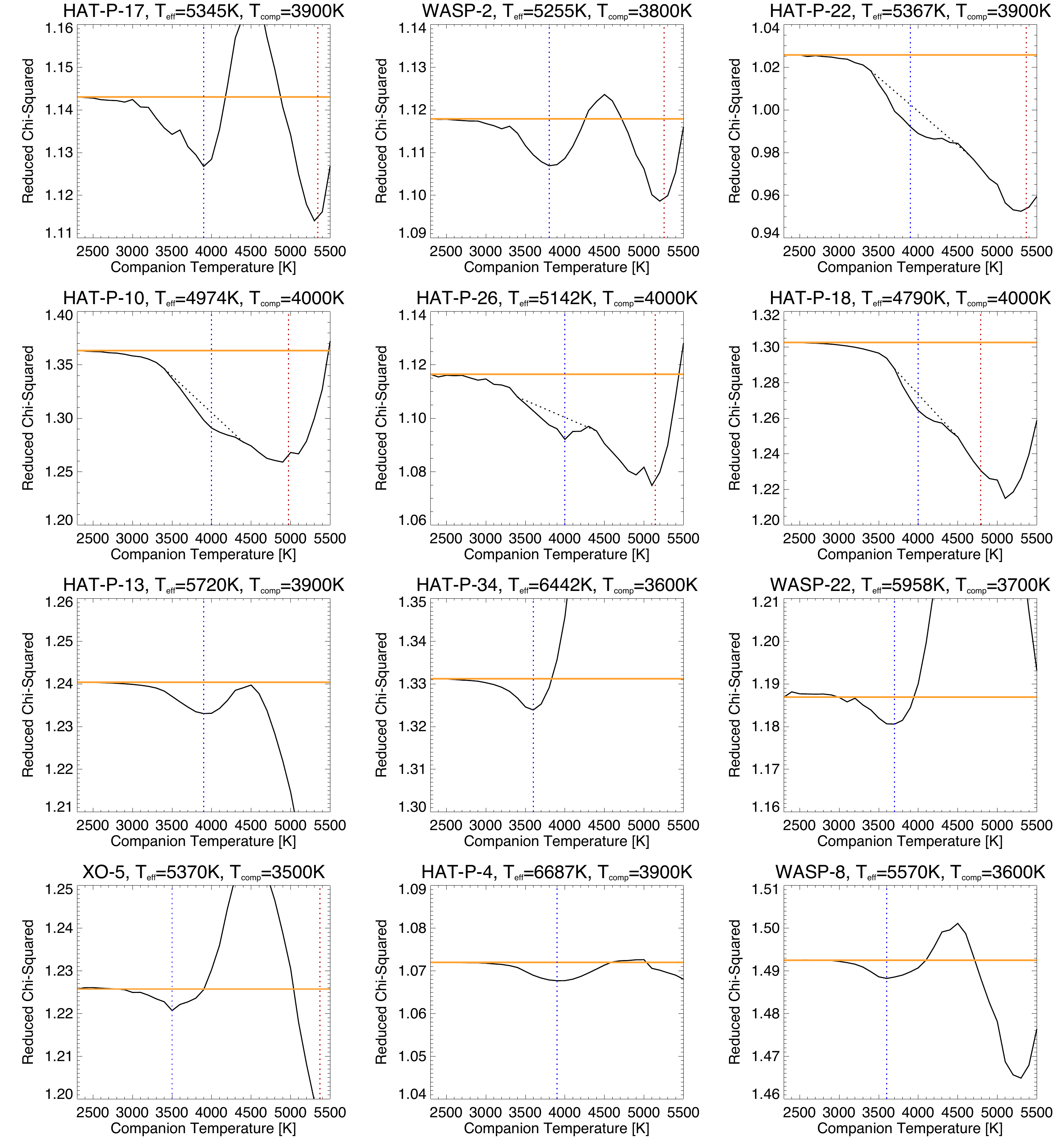}
\caption{Twelve systems with cool candidate companions that pass our detection threshold as given in Table~\ref{detection}. See Fig.~\ref{nondetection} caption for more information. We plot these systems in order of decreasing strength of detection, moving from left to right and top to bottom. HAT-P-17, WASP-2, and WASP-22 were observed twice, and we show the stronger of the two detections here (see Fig~\ref{double} for comparison). We independently resolve the companion to WASP-2 in our AO imaging, as discussed in Section~\ref{WASP-2}.}
\label{detections}
\end{figure*}

There are twelve targets which show a minimum in $\chi^{2}$ that appears to be due to the presence of a cooler stellar companion. We show the results of the model fits for the twelve systems with candidate companions in order of detection strength in Figure~\ref{detections}, and list the corresponding companion temperatures in Table \ref{detection}. The fit results for these twelve systems are shown in Figure~\ref{detections} in order of detection strength. Table~\ref{detection} gives the best-fit companion effective temperature and an upper limit on its projected separation based on the width of the NIRSPEC slit. In all cases, we find that the value of the BIC for the best-fit two-star model is lower than that of the best-fit one-star model for all candidate companions listed in Table~\ref{detection}, thereby justifying the addition of the extra parameter (the temperature of the companion star) to the model.

We present our results in terms of the reduced chi-squared value in order to demonstrate the relative quality of our fits. The reduced chi-squared values are often slightly greater than 1 indicating that our errors are likely underestimated, despite our use of empirical estimates for the measurement errors at each indvidual wavelength as described in Section~\ref{fit}. In addition, the apparent small discontinuity at 5000 K in some of our reduced chi-squared plots is due to a change in the $\lambda_{ref}$ used to calculate the optical depth grid in the PHOENIX models \citep{Husser2013}.   Differences in the shape of the $\chi^{2}$ curve for targets having similar effective temperatures are likely due to different observing conditions. 

\section{Discussion}


\begin{figure}[h]
\centering
\noindent\includegraphics[width=15pc]{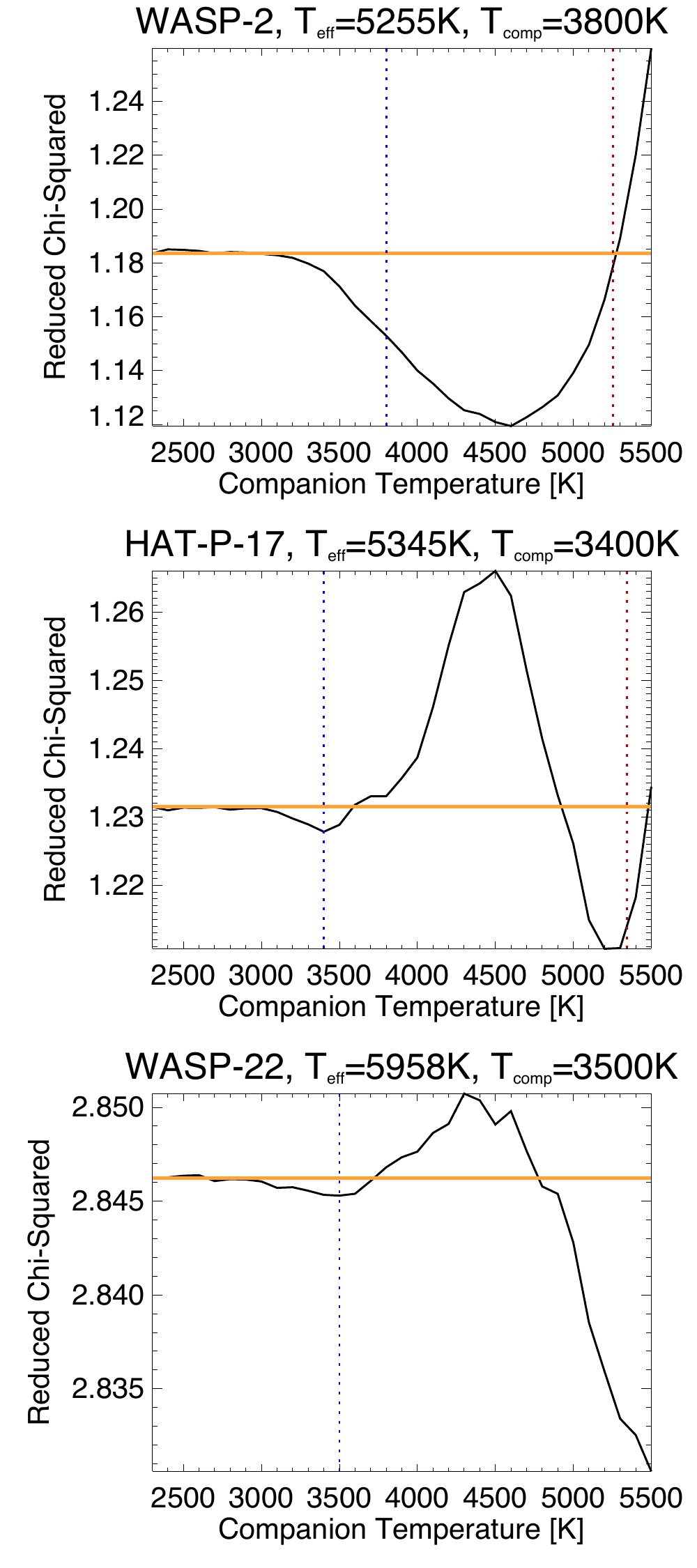}
\caption{A second epoch of data for WASP-2, HAT-P-17, and WASP-22 (see fig.~\ref{detections} caption for more information). For HAT-P-17 and WASP-22 the blue dotted line represents the effective temperature of the best-fit two-star model. For WASP-2, the blue dotted line shows the best-fit two-star model effective temperature suggested by the analysis of the July 2013 data. The companion in this system is resolved in our AO images and has a projected separation of 0.73'', thus it will only be detected in cases where the slit is effectively parallel to the position angle of the two stars. This changing slit orientation may also explain the varying detection strengths for HAT-P-17 and WASP-2.}
\label{double}
\end{figure}

\subsection{Systems with No AO- or RV-Companion Detections}
Here, we list the systems which have a candidate companion detected by the NIRSPEC survey alone.

\subsubsection{HAT-P-17}
As shown in Figure~\ref{detections}, the model fit is significantly improved by the presence of a 3900$^{+200}_{-300}$ K companion in the HAT-P-17 system. According to the size of the NIRSPEC slit, this companion has a projected separation of less than about 36 AU. Our AO survey would only have detected a 3900 K companion if it were outside 60 AU. We also obtained a second spectrum of HAT-P-17 (Fig.~\ref{double}) on a different night in order to make sure that the candidate companion was reliably detected in both data sets. We find consistent results from both nights albeit with varying significance, and report the stronger of the two detections in Table~\ref{detection}. If the projected separation of the companion is comparable to that of the NIRSPEC slit width, the strength of the detection will vary depending on the slit orientation relative to the position angle of the binary.

\subsubsection{HAT-P-26}
For HAT-P-26, we are careful to calculate the improvement in $\Delta\chi^{2}_{red}$ due to the presence of the cooler companion after subtracting off the residual slope due to the primary star, as shown by the dotted line in Figure~\ref{detections}.  We detect a candidate companion in this system having an effective temperature of 4000$^{+100}_{-350}$ K and a projected separation less than 54 AU. A companion having this temperature would not have been detectable by our AO survey.  The shape of our $\chi^{2}_{red}$ curve for this star is analogous to that of the non-detection HAT-P-15 with a 4000 K companion injected into its spectrum as shown in Fig.~\ref{nondetection}. 

\subsubsection{HAT-P-18}
\label{HAT-18}
For the HAT-P-18 system, we detect a candidate companion with an effective temperature of 4000$^{+200}_{-200}$ K and projected separation less than about 66 AU. Our AO survey would only have been sensitive to a 4000K companion if it were outside 140 AU.

 We note that this star is relatively active with \logr=-4.799 \citep{Knutson2010}, and it is therefore likely that the observed minimum is due to spots on the visible face of the star. Previous studies \citep{Frasca2005, Pont2008, Sing2011} have shown that these star spots typically have temperatures 500-1000 K cooler than that of the primary star, in good agreement with the temperature difference observed in this system. 

\subsubsection{HAT-P-34}
In the HAT-P-34 system, we find a 3600$^{+150}_{-250}$ K candidate companion with  a projected separation less than about 103 AU. Such a companion would have been undetectable in our AO survey.

\subsubsection{XO-5}
We detect a candidate companion in the XO-5 system having an effective temperature of 3500$^{+250}_{-150}$ K and a projected separation less than about 104 AU. Our AO survey was unable to detect such a cool companion.

\subsection{Systems with AO-Detected Companions}
We check to see if any of our spectroscopically detected candidate companions are detected independently in the AO imaging survey presented in \cite{Ngo2015}. We obtained our data in parallel with the AO survey, and so did not have any prior knowledge of the positions of resolved stellar companions in these systems that could have been used to determine the optimal slit position angle in the sky. Because NIRSPEC has a slit width of 0.4'', we expect that we are only sensitive to companions with projected separations smaller than 0.2-0.4'' unless they happened to be aligned along the slit in our NIRSPEC observations. We identify three systems with both resolved AO companions and candidate spectroscopic companions, and discuss them each individually below.

\subsubsection{WASP-2}
\label{WASP-2}
An AO companion to WASP-2 was detected first on July 27 2012 and June 22 2013 with an average $T_{\mathrm{eff}}$=3513$\pm$28 K and separation of 0.73''$\pm$0.0015'' \citep{Ngo2015}. Although this projected separation is larger that the NIRSPEC slit width of 0.4'', we expect the companion would still be detectable if it happened to fall along the direction of the slit in our NIRSPEC observations. This appears to be the case, as we detect a stellar companion with $T_{\mathrm{eff}}$=3800$^{+300}_{-350}$ K in our UT 27 August 2012 observation. We also obtained a second spectrum of WASP-2 (Fig.~\ref{double}) on a different night and list the stronger of the two detections in Table~\ref{detection}. Since the projected separation is known to be greater than the size of the NIRSPEC slit, it is likely that varying amounts of companion starlight were gathered on the two nights, producing model fits of differing qualities.

This is the only system in which we independently detect the companion using both spectroscopy and AO imaging, and we therefore use this system to determine an empirical threshold for spectroscopic detections. The $\Delta\chi^{2}_{red}$ for the companion in the system is 0.01, and we therefore adopt a cutoff of 0.005 for determining our list of candidate companions in Table 3. 

\subsubsection{HAT-P-10}
\label{HAT-10}
We detect a bound companion to this star in our \textit{K} band AO imaging with an effective temperature of $T_{\mathrm{eff}}$=3494$\pm$37 K and a projected separation of  0.36''$\pm$0.0015'' \citep{Ngo2015}. Although the projected separation of this companion is smaller than the NIRSPEC slit width, we do not detect it in our spectroscopic observations. As shown in Table~\ref{non} our injection and recovery tests indicate that the companion in this system falls below our detection threshold for this technique. Additionally, if the position angle of the binary companion was perpendicular to the slit and the primary star was located in the middle of the slit, the companion may still have fallen outside the slit aperture.

Although we do not independently recover the AO companion in this system, we do find a minimum in our two-star fits at an effective temperature of 4000$^{+200}_{-200}$ K. Similar to HAT-P-18, HAT-P-10 is relatively active with \logr=-4.82 \citep{Knutson2010}. Therefore, it is likely that the observed chi-squared minimum is due to spots on the visible face of the star. 

\subsubsection{WASP-8}
An AO-companion to WASP-8 was detected on July 27 2012  and August 19 2013 with an average $T_{\mathrm{eff}}$=3591$\pm$157 K and separation of 4.50''$\pm$0.0026'' \citep{Ngo2015}. The separation between WASP-8 and its stellar companion is much larger than the width of the NIRSPEC slit, and it is therefore unlikely that this companion would contribute to our measured NIRSPEC spectrum. Although we identify a weak minimum in our spectroscopic analysis corresponding to a companion with an effective temperature of 3600$^{+350}_{-250}$ K, this minimum falls below our empirical cutoff for a significant detection.

\subsection{Systems with RV-Detected Companions}
Here, we ask whether or not the candidate spectroscopic companions could have caused the RV trends. We therefore consider whether or not any of the candidate spectroscopic stellar companions detected in this study might be responsible for the radial velocity accelerations reported in \cite{Knutson2014}. 

For the systems where the mass of NIRSPEC candidate companion is consistent with the measured radial velocity trend, we calculate the system's likely angle from face-on given by 
\begin{equation}
\mathrm{sin}\theta=\frac{\dot{\gamma}a_{c}^{2}}{GM_{c}}
\label{semiamplitude}
\end{equation}
as in Winn et al. (2009), where $\dot{\gamma}$ is the RV trend measured by \cite{Knutson2014}, $M_{c}$ is the mass of the NIRSPEC candidate companion calculated from $T_{\mathrm{eff}}$ according to \cite{Baraffe2003}, and $a_{c}$ is the candidate companion's semi-major axis. This latter value is the least well-known and only vaguely constrained by the size of the NIRSPEC slit. 


\subsubsection{HAT-P-22}
In \cite{Knutson2014} we reported a radial velocity acceleration due to a companion with $M\mathrm{sin}i$  between 0.7-125 $M_{\mathrm{Jup}}$ and semi-major axis of 3.0-28 AU (1$\sigma$ constraints), where upper limits on the companion mass and orbit were calculated based on the AO non-detection. We report a candidate spectroscopic companion in this system with an effective temperature of 4000$^{+250}_{-400}$ K, corresponding to a mass of 660$^{+75}_{-175}$ $M_{\mathrm{Jup}}$ and a maximum projected separation of 33 AU (0.4''). According to equation~\ref{semiamplitude}, if this companion has an orbital semi-major axis less than 33 AU it must have a face-on orbit in order to be consistent with the observed RV trend. It is also possible that the companion is located at larger semi-major axes, but was observed at a time when it had a relatively small projected separation and/or small radial velocity slope.

\subsubsection{HAT-P-10}
In \cite{Knutson2014} we detected a long-term radial velocity acceleration in the HAT-P-10 system, which was consistent with having been caused by a directly imaged AO companion reported in \cite{Ngo2015}. As discussed in Section~\ref{HAT-10}, the NIRSPEC detection is likely an indication of stellar activity. 

\subsubsection{HAT-P-13}
HAT-P-13 has two companions detected with RV. The first, HAT-P-13c, has an $M\mathrm{sin}i$ of 14.23-15.18$M_{\mathrm{Jup}}$ and a semi-major axis of 1.24-1.28 AU (1$\sigma$ constraints). HAT-P-13d has an $M\mathrm{sin}i$  of 15-200 $M_{\mathrm{Jup}}$ and a semi-major axis of 12-37 AU (1$\sigma$ constraints). Our candidate spectroscopic companion has an effective temperature of 3900$^{+300}_{-350}$ K, which corresponds to a mass of 0.602$^{+0.086}_{-0.179}$ $M_{\odot}$ or 630$^{+91}_{-187}$ $M_{\mathrm{Jup}}$, and projected separation $\lesssim$ 85.6 AU. If the candidate spectroscopic companion were HAT-P-13d identified by our RV survey, then it must have an inclination within 5$^{\circ}$ of face-on. However, \cite{Winn2010} argue that this system is likely coplanar, as otherwise the influence of the outer companions would tend to misalign the orbit of the inner transiting hot Jupiter with respect to the star's spin axis. They find that the innermost planet's orbit is well-aligned with the star's spin axis, suggesting that the  $M\mathrm{sin}i$  values of the outer two companions are likely close to their true masses. For the same reason we argue here that any outer stellar companion must also be coplanar with the orbits of the planets in this system. This constraint might be relaxed if the stellar companion was distant enough that Kozai-type oscillations would not occur (see Ngo et al. 2015 and references therein), but this would require that the system was observed at a time when the projected separation between the companion and the primary was small in order to remain consistent with both our spectroscopic detection and our non-detection in AO images of this system. If we require the companion to be coplanar with the inner planets, then our radial velocity measurements allow us to rule out scenarios where the stellar companion is located interior to 40 AU on a high inclination orbit.

\subsubsection{WASP-22}
We detect a radial velocity acceleration in this system corresponding to a companion with $M\mathrm{sin}i$ between 7-500 $M_{\mathrm{Jup}}$ and a semi-major axis between 6-40 AU. The candidate spectroscopic companion in this system has an effective temperature of 3700$^{+150}_{-300}$ K, which corresponds to a mass of 0.523$^{+0.063}_{-0.253}$ $M_{\odot}$ or 548$^{+66}_{-266}$ $M_{Jup}$, and separation $\lesssim$ 120 AU. We therefore conclude that our spectroscopic candidate could have caused the radial velocity acceleration measured in this system if it is on an orbit within 10$^{\circ}$ of face-on. We also obtained a second spectrum of WASP-22 (Fig.~\ref{double}) on a different night in order to make sure that the candidate companion was reliably detecetd in both data sets. We find consistent results from both nights albeit with varying significance, and report the stronger of the two detections in Table~\ref{detection}.

\subsection{Detections below Empirical Threshold for Significance}
Here we discuss the results for the systems showing marginal detections of a companion star in order to determine the effectiveness of our threshold of $\Delta\chi^{2}_{red} \ge$ 0.005.

\subsubsection{HAT-P-4}
The radial velocity companion in this system is constrained to have $M\mathrm{sin}i$ between 1.5-310 $M_{\mathrm{Jup}}$ and a semi-major axis of 5-60 AU (1$\sigma$ constraints). We identify a marginally significant spectroscopic signal corresponding to a stellar companion with an effective temperature of 3900$^{+450}_{-400}$ K and a mass of 0.602$^{+0.123}_{-0.224}$ $M_{\odot}$ or 631$^{+129}_{-236}$ $M_{Jup}$ at a projected separation of less than 120 AU. If the RV signal is caused by this companion, then it must either be located in a short-period orbit within 4$^{\circ}$ of face-on, or on a more distant orbit with a small projected separation and/or small radial velocity slope.

\subsubsection{WASP-8}
The radial velocity acceleration in this system displays significant curvature, and we are therefore able to place relatively tight constraints on the mass and orbital separation of the companion responsible for the acceleration. In this case we find that the companion has $M\mathrm{sin}i$ between 6.3-10.7 $M_{\mathrm{Jup}}$, and we therefore conclude that it is most likely a low-mass brown dwarf or planetary companion. Our candidate spectroscopic companion in this system is a relatively weak detection and has an effective temperature of 3600$^{+350}_{-250}$ K and a projected separation of less than 35 AU. Therefore, if this NIRSPEC companion candidate is in fact a true companion, then the NIRSPEC companion is not the same as the RV companion. We also detect an AO companion in this system with a temperature of 3590 K and a projected separation of 4.50'', which is too large to be detected in our NIRSPEC observation. It seems unlikely that this system would contain a hot Jupiter, an outer planetary or brown dwarf companion, and two stellar companions with widely varying orbital separations, and we therefore conclude that the NIRSPEC detection in this system is unlikely to be real. This would not be surprising, as this is the weakest of the candidate companion detections listed in Table~\ref{detection}. \newline

Given the specious nature of the candidate companion in the WASP-8 system, we assert our empirical detection theshold of $\Delta\chi^{2}_{red} \ge$ 0.005 is a reasonable lower limit for identifying candidate companions in these systems.

\section{False Detections Due to Star Spots}
We identify the spectroscopic signal of star spots in the spectra of HAT-P-18 and HAT-P-10, as discussed in sections~\ref{HAT-18} and~\ref{HAT-10}. Converting the area ratio for each star and its ``candidate companion" suggests that the fraction of the stellar surface covered by star spots is 39\% and 37\% for HAT-P-18 and HAT-P-10, respectively. This level of star spot coverage is somewhat high, although not unheard of (e.g. Jackson and Jeffries 2013). In an attempt to gain more specific information on the fraction of these stars covered by star spots, we vary the contribution of the cool star spectrum to the two-star model. However, we find that there is no clear minimum separate from that of the stellar effective temperature. This suggests that the temperature of the star spots is degenerate with their fractional area over the range of effective temperatures considered in our fits.

\begin{figure}[h]
\centering
\noindent\includegraphics[width=20pc]{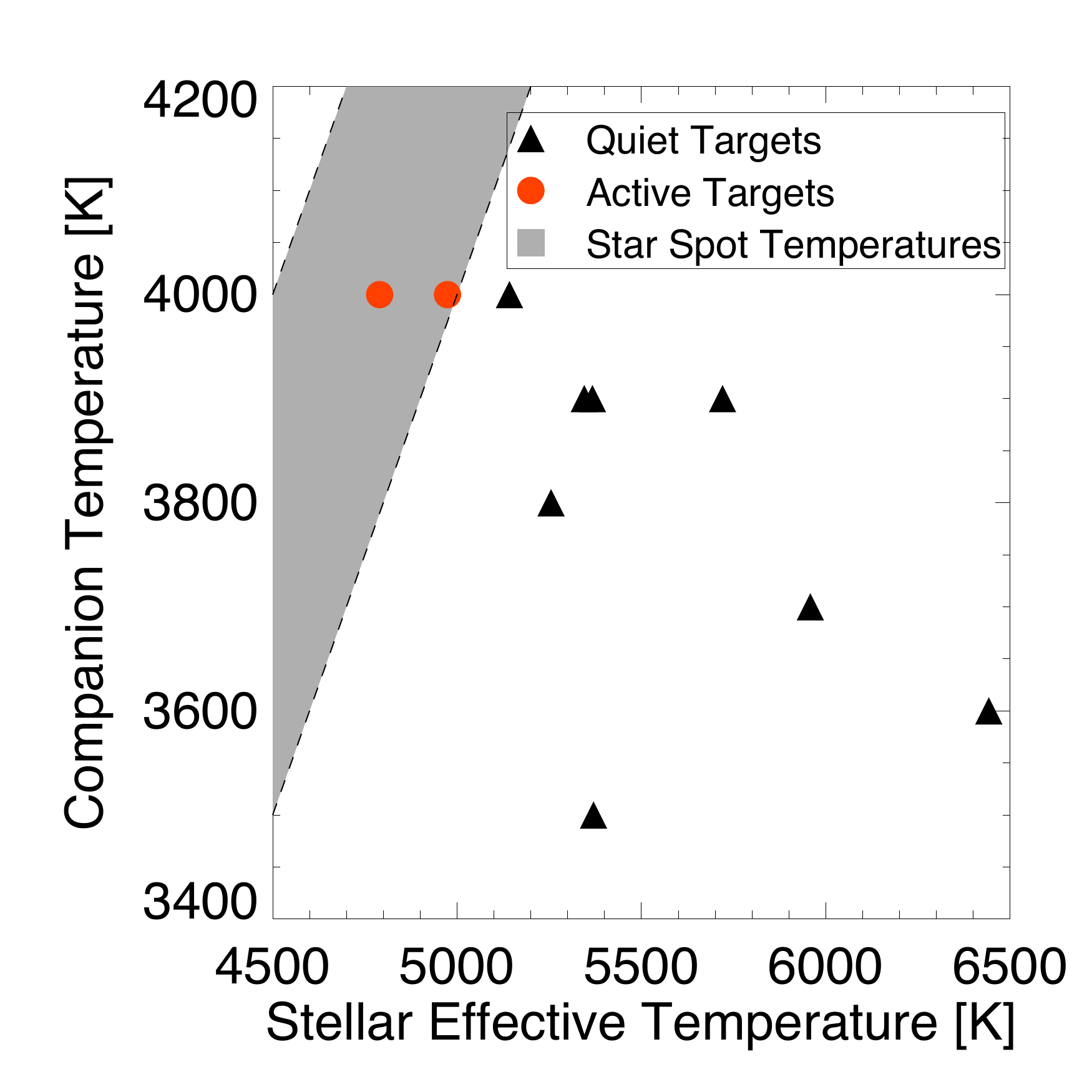}
\caption{ Effective temperatures of the twelve targets and their candidate companions. Targets having  \logr $<$ -4.9 are classified as quiet stars and plotted as black filled triangles.  HAT-P-10 and HAT-P-18 have log(R') $>$ -4.9 indicating moderate levels of activity, and are shown as red filled circles in this plot. The shaded region represents the expected star spot temperatures as a function of stellar effective temperature according to \cite{Frasca2005}, \cite{Pont2008}, and \cite{Sing2011}. }
\label{activity}
\end{figure}

We next consider the candidate companions around the other stars in our sample.  If we attribute these spectroscopic signals to star spots we find that the fractional flux contributions of the candidate companions correspond to star spot coverage fractions between 9-37\%. Although we cannot distinguish between star spots and low mass companions on the basis of our spectra alone, we consider it unlikely that all of the candidate companions presented in this paper are in fact due to stellar activity.  HAT-P-18 and HAT-P-10 are relatively active stars with relatively low effective temperatures. As shown in Figure~\ref{activity}, the remaining systems with candidate companions appear to be relatively quiet stars as measured by \logr, with the caveat that this index may not be a reliable activity indicator for stars with effective temperatures greater than 6000-6200 K (e.g., Knutson et al. 2010). Furthermore, the candidate companion temperatures implied by our fits to these quiet stars are much cooler than would be expected for star spots.  We therefore conclude that HAT-P-10 and HAT-P-18 are the only systems in which our detection of a candidate companion can plausibly be explained as stellar activity.

\section{Companion Fraction}
Although our candidate spectroscopic companions still require additional confirmation, we can nonetheless calculate an upper limit to the companion fraction in our sample corresponding to the case where all candidates are confirmed as real. These companions have $T_\mathrm{eff}$ = 3500-4000 K and projected separations less than 125 AU. We exclude the seven cool stars listed in Section 3 from this calculation, as our NIRSPEC observations are not sensitive to low-mass stellar companions in these systems. Of the systems with candidate companion detections, we exclude HAT-P-10  and HAT-P-18  as the detections in these systems are likely due to stellar activity.  We also exclude HAT-P-4 and WASP-8 as the detections in these systems fall below our empirical threshold for significance, and WASP-2 as AO imaging indicates that the companion in this system has a projected separation greater than 0.4".  We find an \textit{uncorrected} binary fraction of 18\%  $\pm$ 6\% for the full survey sample,  20\%  $\pm$ 9\% for the subset of short period gas giant planets with eccentric and/or misaligned orbits, and 25\% $\pm$ 9\% for the subset of planets with apparently well-aligned and circular orbits. 

In order to correct for survey incompleteness we must make some assumptions about the properties of the underlying population of stellar companions in our target sample.  We consider two scenarios, including one where the detected companions are all on wide orbits similar to those of the companions detected in our previous AO survey (Ngo et al. 2015), but happen to be observed at a time when they have a small projected separation.  We also consider a scenario in which the companions are on orbits that are too close in to have been detected by our AO survey but far enough out to be consistent with a non-detection in our radial velocity survey (Knutson et al. 2014). For each underlying population, we calculate our sensitivity $S_i$ for each survey target $i$ to companions having effective temperatures within the ranges shown in Table~\ref{non} and projected separations less than 0.4'' at the time of observation. We compute the likelihood $L$ of obtaining our specific observations as follows:
\begin{equation}
L(\eta)=\prod\limits_{i=1}^{N}(S_i \eta)^{d_i}(1-S_i \eta)^{(1-d_i)}
\label{likelihood}
\end{equation}
where $\eta$ is the true/intrinsic companion fraction and $d_i$=1 for systems with NIRSPEC detections and $d_i=0$ for systems without NIRSPEC detections.

\subsection{Are these companions drawn from the population of wide companions detected in previous surveys? }
In this scenario we assume that the observed NIRSPEC companions are part of the same population as the wide-separation companions detected in our AO survey.  In this case the NIRSPEC companions would be located on relatively wide orbits, but happen to be observed at a time when they have projected separations of less than 0.4". We calculate our sensitivity $S_i$ to this population for each target star by generating 10$^6$ simulated binary companions with orbital elements and mass ratio distributions drawn from field star surveys \citep{Raghavan2010} and periods between 10$^4$ and 10$^{7.5}$ days, as described in \cite{Ngo2015}.  We then determine the fraction of these simulated companions that would have been detected by our NIRSPEC observations. 

For this population, we find that our average survey sensitivity is $28\%$ with a standard deviation of $9\%$.  If we assume an intrinsic companion fraction of $49\% \pm 9\%$ as reported in Ngo et al. (2015), then we would expect to detect companions in $13\% \pm 5\%$ of the systems in our NIRSPEC sample. This is entirely consistent with our actual raw companion fraction of $18\%  \pm 6\%$, which is in fact an upper limit predicated on the assumption that all of the candidate companions are real.  We therefore conclude that our data do not require these systems to have an additional population of close-in stellar companions, but are instead consistent with being drawn from the same population of wide-separation companions detected in our previous AO survey.

Using equation 5 and assuming that the underlying population has the same mass ratio and period distribution as the AO sample, we find that the corrected companion fraction is  $\eta_T = 58\% \pm 19\%$, consistent with the  $49\% \pm 9\%$ companion fraction from our AO survey.  The corrected companion fraction for the subset of planets with misaligned and/or eccentric orbits is $\eta_M = 51\% \pm 30\%$, and for well-aligned planets on circular orbits is $\eta_C = 65\% \pm 30\%$.  These occurrence rates are consistent with one another and suggest that there is no obvious preference for misaligned hot Jupiter systems to have stellar companions drawn from the AO sample. 

\subsection{Are these companions drawn from the population of intermediate companions detected in field star surveys?}
We next consider whether or not the candidate companions detected in this survey might be a field star population of intermediate-period companions located too close in to be detectable in AO images and too far out to be detected with with radial velocity monitoring.  We calculate the sensitivity of our survey to this population by simulating $10^6$ binary companions with orbital elements and mass ratio distributions drawn from \cite{Raghavan2010}.  We only consider systems with periods less than $10^4$ days (i.e., interior to our cutoff for the AO survey) and those which do not create detectible RV trends. In our formulation, a binary system has no detectable RV trend if the following criterion is satisfied:
\begin{equation}
\frac{GM_{sim}\mathrm{sin}\theta}{a_{sim}^2}<\dot{\gamma}_{i} + 3\sigma_{\dot{\gamma}_i}
\label{likelihood}
\end{equation}
where $M_{sim}$ and $a_{sim}$ are the mass and semi-major axis of the simulated binary companion and $\theta$ is the angle from face-on, as in equation ~\ref{semiamplitude}. $\dot{\gamma}_{i}$ is the RV trend slope measured for system $i$ in ~\cite{Knutson2014} and $\sigma_{\dot{\gamma}_i}$ is the error on that measurement. 

For this population, we find that our average survey sensitivity is $49\%$ with a standard deviation of $14\%$. This corresponds to a completeness-corrected companion fraction  $\eta_T = 34\% \pm 11\%$ for the total sample, $\eta_M = 29\% \pm 17\%$ for the misaligned sample, and $\eta_C = 38\% \pm 17\%$ for the control sample.  As before, we conclude that there is no detectable correlation between the presence of a stellar companion and the orbital properties of the transiting gas giant planet. We find that stellar companions that meet these two criterion typically have semi-major axes between 0.4 - 12.1 AU. Of field stars surveyed by \cite{Raghavan2010},  $17\% \pm 2\%$ had companions in this semi-major axis range. We therefore conclude that, if these companions are drawn from a unique imtermediate period sample, our upper limit on the companion fraction for this population is consistent with the rates for field stars with comparable separations.

\section{Conclusion}
We perform a spectroscopic search of fifty hot Jupiter host stars in order to search for blended lines due to cool stellar companions with projected separations of less than 0.4''. We detect eight candidate companions having effective temperatures ranging from 3500-4000 K. This method is complementary to our previous AO imaging survey of these same systems, which was sensitive to companions with projected separations larger than 0.4-0.5''. It also allows us to determine whether or not any of the radial velocity trends in these systems might be due to high inclination stellar binaries as opposed to planetary or brown dwarf companions. Our detection sensitivity and corrected companion fractions are consistent with a scenario in which all of the observed companions are located at larger orbital separations consistent with the population of companions detected in our previous AO survey.  Our results are also consistent with a scenario in which the observed companions are located at intermediate separations (0.4-12 AU) with a frequency comparable to that of field stars. 

Regardless of the underlying population model, we find no evidence for a correlation between the presence of a spectroscopic stellar companions and the spin-orbit misalignment or orbital eccentricity of the transiting gas giant planets in these systems. Other surveys have suggested that misaligned hot Jupiters preferentially orbit stars having \teff $>$ 6200 K \citep{Winn2010}. We note that hot stars also have higher intrinsic binary fractions (e.g., Duchene \& Kraus 2013), which might lead to a spurious correlation between spin-orbit alignment and stellar multiplicity if most of our misaligned planets are located preferentially around hot stars. We selected our sample of well-aligned planets on circular orbits to have approximately the same distribution of stellar masses as our sample of misaligned and/or eccentric planets (see section 2.1 of \cite{Knutson2014}) in order to ensure that this was not an issue.  We also note that this study identifies only one system with \teff $>$ 6200 K containing a candidate companion, providing additional confirmation that our results are unaffected by this correlation. Other studies have suggested that tides can remove non-zero obliquities for planets orbiting stars cooler than 6200 K \citep{Spalding2015}, although planets orbiting hotter stars with less efficient tides should still retain their primordial spin-orbit alignments. In Ngo et al. (2015) we addressed this issue by repeating our spin-orbit alignment correlation test for the subset of stars hotter than 6200 K, and found no evidence for a correlation between spin-orbit alignment and stellar multiplicity.  Although we could in theory repeat this analysis with our new sample of candidate companions, in practice we are limited in this case by the small number of companions detected in this survey. If misaligned orbits are instead the results of a primordial misalignment in the protoplanetary disk \citep{Spalding2014}, the stellar companions responsible for the misalignment might have been lost to dissolution or exchange in the cluster, therefore weakening the observed correlation between spin-orbit misalignment and the presence of a stellar companion at the current epoch. Finally, there would be no correlation between companion stars and misaligned planets if misalignment were caused by magnetic torque \citep{Lai2011} or turbulent accretion \citep{Bate2010, Fielding2014}, neither of which require the presence of a stellar companion. Although our data appear to be consistent with this hypothesis, these models are unable to reproduce the observed obliquity distribution for short-period gas giant planets \citep{Spalding2014b}.

An independent confirmation of these spectroscopic detections would allow us to reliably calculate the statistical significance of our spectroscopic detections and to combine the results of this study with our previous RV and AO surveys in order to provide an improved estimate of the stellar multiplicity rate for these systems. As shown in \cite{Teske2015}, combined direct imaging and spectroscopic surveys provide an unparalleled look at the properties of stellar companions in these systems. In the future we plan to obtain high-contrast imaging follow-up observations using angular differential imaging on NIRC2 at Keck \citep{Marois2006} and the Differential Speckle Survey Instrument at Gemini Observatory \citep{Horch2009}, which will achieve better constrasts at small separations than our previous \textit{K} band AO images with NIRC2. Because our candidate spectroscopic companions have negligible radial velocity offsets from the primary and are not typically detected by long-term radial velocity monitoring of the host star, we conclude that they are unlikely to have very small orbital separations and therefore should be resolvable with improved high-contrast imaging. In systems where we are able to directly resolve the candidate spectroscopic stellar companions, we will be able to place improved constraints on their masses and orbits. 

\acknowledgments{The authors would like to thank John Johnson for helpful discussions contributing to this paper. The authors wish to recognize and acknowledge the very significant cultural role and reverence that the summit of Mauna Kea has always had within the indigenous Hawaiian community.  We are most fortunate to have the opportunity to conduct observations from this mountain. The data presented herein were obtained at the W.M. Keck Observatory, which is operated as a scientific partnership among the California Institute of Technology, the University of California and the National Aeronautics and Space Administration. The Observatory was made possible by the generous financial support of the W.M. Keck Foundation.}


\begin{thebibliography}{79}
\providecommand{\natexlab}[1]{#1}
\providecommand{\url}[1]{\texttt{#1}}
\expandafter\ifx\csname urlstyle\endcsname\relax
  \providecommand{\doi}[1]{doi: #1}\else
  \providecommand{\doi}{doi: \begingroup \urlstyle{rm}\Url}\fi

\bibitem[Adams et~al. (2012)]{Adams2012}
Adams, E.R., Ciardi, D.R., Dupree, A.K. et al. 2012, AJ, 144, 42

\bibitem[Albrecht et~al.(2012)Albrecht, Winn, Johnson, Howard, Marcy, Butler,Arriagada, Crane, Shectman, Thompson, Hirano, Bakos, and Hartman]{Albrecht2012}
Albrecht, S., Winn, J. N., Johnson, J. A., et al. 2012b, ApJ, 757, 18

\bibitem[Anderson et~al.(2010)Anderson, Hellier, Gillon, Triaud, Smalley, Hebb, Cameron, Maxted, Queloz, West, Bentley, Enoch, Horne, Lister, Mayor, Parley, Pepe, Pollacco, S\'{e}gransan, Udry, and Wilson]{Anderson2010}
Anderson, D. R., Hellier, C., Gillon, M., et al. 2010, ApJ, 709, 159

\bibitem[Anderson et~al.(2011)Anderson, {Collier Cameron}, Gillon, Hellier, Jehin, Lendl, Queloz, Smalley, Triaud, and Vanhuysse]{Anderson2011}
Anderson, D. R., Collier Cameron, A., Gillon, M., et al. 2011, A\&A, 534, A16

\bibitem[Bakos et~al.(2009)Bakos, P\'{a}l, Torres, Sipőcz, Latham, Noyes, Kov\'{a}cs, Hartman, Esquerdo, Fischer, Johnson, Marcy, Butler, Howard, Sasselov, Kov\'{a}cs, Stefanik, L\'{a}z\'{a}r, Papp, and S\'{a}ri]{Bakos2009}
Bakos, G. A.,P\'{a}l, A., Torres, G., et al. 2009b, ApJ, 696, 1950

\bibitem[Bakos et~al.(2010)Bakos, Torres, P\'{a}l, Hartman, Kov\'{a}cs, Noyes, Latham, Sasselov, Sipőcz, Esquerdo, Fischer, Johnson, Marcy, Butler, Isaacson, Howard, Vogt, Kov\'{a}cs, Fernandez, Mo\'{o}r, Stefanik, L\'{a}z\'{a}r, Papp, and S\'{a}ri]{Bakos2010}
Bakos, G. A., Torres, G.,P\'{a}l, A., et al. 2010, ApJ, 710, 1724

\bibitem[Bakos et~al.(2011)Bakos, Hartman, Torres, Latham, Kov\'{a}cs, Noyes, Fischer, Johnson, Marcy, Howard, Kipping, Esquerdo, Shporer, B\'{e}ky, Buchhave, Perumpilly, Everett, Sasselov, Stefanik, L\'{a}z\'{a}r, Papp, and S\'{a}ri]{Bakos2011}
Bakos, G.~\'{A}., J.~Hartman, G.~Torres, et al. 2011, ApJ, 742, 116

\bibitem[Bakos et~al.(2012)Bakos, Hartman, and Torres]{Bakos2012}
Bakos, G\'{A} , JD~Hartman, G~Torres, et al. 2012, AJ, 144, 19

\bibitem[Baraffe et~al.(2003)Baraffe, Chabrier, Barman, Allard, and Hauschildt]{Baraffe2003}
Baraffe, I., Chabrier, G., Allard, F., \& Hauschildt, P. H. 1998, A\&A, 337, 403

\bibitem[Barclay et~al.(2012)Barclay, Huber, Rowe, Fortney, Morley, Quintana, Fabrycky, Barentsen, Bloemen, Christiansen, Demory, Fulton, Jenkins, Mullally, Ragozzine, Seader, Shporer, Tenenbaum, and Thompson]{Barclay2012}
Barclay, T., Huber, D., Rowe, J. F., et al. 2012, ApJ, 761, 53

\bibitem[Bardalez Gagliuffi et~al.(2014)]{Gagliuffi2014}
Bardalez Gagliuffi, D. C., Burgasser, A. J., Gelino, C. R., et al. 2014, AJ, 794, 143

\bibitem[Bate et~al.(2010)]{Bate2010}
Bate, M. R., Lodato, G., \& Pringle, J. E. 2010, MNRAS, 401, 1505

\bibitem[Batygin et al.(2011)]{Batygin2011}
Batygin, K., Morbidelli, A., \& Tsiganis, K. 2011, A\&A, 533, A7

\bibitem[Batygin(2012)]{Batygin2012}
Batygin, K. 2012, Nature, 491, 418

\bibitem[Beaug\'e \& Nesvorn\'y(2012)]{Beauge2012}
Beaug\'e, C., \& Nesvorn\'y, D. 2012, ApJ, 751, 119

\bibitem[Bechter et~al.(2014)]{Bechter2014}
Bechter, E. B., Crepp, J. R., \& Ngo, H. et al. 2014, ApJ, 788, 2

\bibitem[Bergfors et~al.(2013)Bergfors, Brandner, Daemgen, Biller, Hippler, Janson, Kudryavtseva, Geissler, Henning, and Kohler]{Bergfors2012}
Bergfors, C., Brandner, W., Daemgen, S., et al. 2013, MNRAS, 428, 182

\bibitem[Birkby et al. (2013)]{Birkby2013}
Birkby, J. L., de Kok, R. J., \& Brogi, M. et al. 2013, MNRAS, 436, L35

\bibitem[Bodenheimer et~al.(2000)]{Bodenheimer2000}
Bodenheimer, P., Hubickyj, O., \& Lissauer, J. J. 2000, Icarus, 143, 2

\bibitem[Bonfils et~al.(2005)Bonfils, Delfosse, Udry, Santos, and Forveille]{Bonfils2005}
Bonfils, X., Delfosse, X., Udry, S., et al. 2005, A\&A, 442, 635

\bibitem[Bonnell et~al.(2001)]{Bonnell2001}
Bonnell, I.A., Smith, K.W., Davies, M.B. et al. 2001, MNRAS, 322, 859

\bibitem[Boogert et~al.(2002)]{Boogert2002}
Boogert, A. C. A., Blake, G. A., \& Tielens, A. G. G. M. 2002, ApJ, 577, 271

\bibitem[Brown et~al.(2012)Brown, {Collier Cameron}, D\'{\i}az, Doyle, Gillon, Lendl, Smalley, Triaud, Anderson, Enoch, Hellier, Maxted, Miller, Pollacco, Queloz, Boisse, and H\'{e}brard]{Brown2012}
Brown, D. J. A., Collier Cameron, A., Diaz, R. F., et al. 2012b, ApJ, 760, 139

\bibitem[Buchhave et~al.(2010)Buchhave, Bakos, Hartman, Torres, Kov\'{a}cs, Latham, Noyes, Esquerdo, Everett, Howard, Marcy, Fischer, Johnson, Andersen, Fűr\'{e}sz, Perumpilly, Sasselov, Stefanik, B\'{e}ky, L\'{a}z\'{a}r, Papp, and S\'{a}ri]{Buchhave2010}
Buchhave, L. A., Bakos, G.~\'{A}., Hartman, J. D., et al. 2010, ´ ApJ, 720, 1118

\bibitem[Buchhave et~al.(2011)Buchhave, Bakos, Hartman, Torres, Latham, Andersen, Kov\'{a}cs, Noyes, Shporer, Esquerdo, Fischer, Johnson, Marcy, Howard, B\'{e}ky, Sasselov, Fűr\'{e}sz, Quinn, Stefanik, Szklen\'{a}r, Berlind, Calkins, L\'{a}z\'{a}r, Papp, and S\'{a}ri]{Buchhave2011}
Buchhave, L. A., Bakos, G.~\'{A}, Hartman, J. D., et al. 2011, ´ ApJ, 733, 116

\bibitem[Burgasser et~al.(2010a)]{Burgasser2010a}
Burgasser, A. J., Cruz, K. L., Cushing, M., et al. 2010a, ApJ, 710, 1142

\bibitem[Burke et~al.(2007)Burke, McCullough, Valenti, Johns‐Krull, Janes, Heasley, Summers, Stys, Bissinger, Fleenor, Foote, Garcia‐Melendo, Gary, Howell, Mallia, Masi, Taylor, and Vanmunster]{Burke2007}
Burke, C. J., McCullough, P. R., Valenti, J. A., et al. 2007, ApJ, 671, 2115

\bibitem[Carter et~al.(2009)Carter, Winn, Gilliland, and Holman]{Carter2009}
Carter, J. A., Winn, J. N., Gilliland, R., \& Holman, M. J. 2009, ApJ, 696, 241

\bibitem[Cheetham et~al.(2015)]{Cheetham2015}
Cheetham, A.C., Kraus, A.L., Ireland, M.J., et al. 2014, ApJ(accepted), arXiv:1509.05217

\bibitem[Christian et~al.(2009)Christian, Gibson, Simpson, Street, Skillen, Pollacco, {Collier Cameron}, Joshi, Keenan, Stempels, Haswell, Horne, Anderson, Bentley, Bouchy, Clarkson, Enoch, Hebb, H\'{e}brard, Hellier, Irwin, Kane, Lister, Loeillet, Maxted, Mayor, McDonald, Moutou, Norton, Parley, Pont, Queloz, Ryans, Smalley, Smith, Todd, Udry, West, Wheatley, and Wilson]{Christian2009}
Christian, D. J., Gibson, N. P., Simpson, E. K., et al. 2009, MNRAS, 392, 1585

\bibitem[Claret(2000)]{Claret2000}
Claret, A~. 2010, A\&A, 529, A79

\bibitem[Dawson et~al.(2013)]{Dawson2013}
Dawson, R.I., \& Murray-Clay, R.A. 2013, ApJ, 767, L24

\bibitem[de Mooij et al. (2012)]{deMooij2012}
de Mooij, E.J.W., Brogi, M., de Kok, R.J., et al. A\&A 538, A46 (2012)

\bibitem[Doyle et~al.(2011)]{Doyle2011}
Doyle, L.R., Carter, J.A., Fabrycky, D.C., et al. 2011, Science, 333, 1602

\bibitem[Doyle et~al.(2013)Doyle, Smalley, Maxted, Anderson, Cameron, Gillon, Hellier, Pollacco, Queloz, Triaud, and West]{Doyle2012}
Doyle, A. P., Smalley, B., Maxted, P. F. L., et al. 2013, MNRAS, 428, 3164

\bibitem[Duchene \& Kraus(2013)]{Duchene2013}
Duchene, G., \& Kraus, A. 2013, ARA\&A, 51, 269

\bibitem[Duquennoy \& Mayor(1991)]{Duquennoy1991}
Duquennoy, A., \& Mayor, M. 1991, A\&A, 248, 485

\bibitem[Eggenberger et~al.(2007)Eggenberger, Udry, Chauvin, and Mayor]{Eggenberger2007}
Eggenberger, A., Udry, S., Chauvin, G., et al. 2007, A\&A, 474, 273

\bibitem[Fabrycky \& Tremaine(2007)]{Fabrycky2007}
Fabrycky, D., \& Tremaine, S. 2007, ApJ, 669, 1298

\bibitem[Fielding et~al.(2014)]{Fielding2014}
Fielding, D. B., McKee, C. F., Socrates, A., Cunningham, A. J., \& Klein, R. I. 2014, MNRAS, 450, 3

\bibitem[Frasca et~al.(2005)Frasca, Biazzo, Catalano, Marilli, Messina, and Rodon\`{o}]{Frasca2005}
Frasca, A., K.~Biazzo, S.~Catalano, et al.  2005, A\&A, 432, 2

\bibitem[Gibson et~al.(2008)Gibson, Pollacco, Simpson, Joshi, Todd, Benn, Christian, Hrudkov\'{a}, Keenan, Meaburn, Skillen, and Steele]{Gibson2008}
Gibson, N. P., Pollacco, D., Simpson, E. K., et al. 2008, A\&A, 492, 603

\bibitem[Goldreich \& Tremaine(1980)]{Goldreich980}
Goldreich, P., \& Tremaine, S. 1980, ApJ, 241, 425

\bibitem[Gray(2005)]{Gray2005}
Gray, D.F. 2005, \newblock \emph{{The Observation and Analysis of Stellar Photospheres}}. \newblock Cambridge University Press, Cambridge, 3 edition

\bibitem[Guenther et~al.(2013)]{Guenther2013}
Guenther, E.W., Fridlund, M., Alonso, R. et al. 2013, A\&A, 556, 75

\bibitem[Hao et~al.(2013)]{Hao2013}
Hao, W., Kouwenhoven, M.B.N., \& Spurzem, R. 2013, MNRAS, 433, 867

\bibitem[Hartman et~al.(2009)Hartman, Bakos, Torres, Kov\'{a}cs, Noyes, P\'{a}l, Latham, Sipőcz, Fischer, Johnson, Marcy, Butler, Howard, Esquerdo, Sasselov, Kov\'{a}cs, Stefanik, Fernandez, L\'{a}z\'{a}r, Papp, and S\'{a}ri]{Hartman2009}
Hartman, J. D., Bakos, G.~\'{A}., Torres, G., et al. 2009, ApJ, 706, 785

\bibitem[Hartman et~al.(2011)Hartman, Bakos, Torres, Latham, Kov\'{a}cs, B\'{e}ky, Quinn, Mazeh, Shporer, Marcy, Howard, Fischer, Johnson, Esquerdo, Noyes, Sasselov, Stefanik, Fernandez, Szklen\'{a}r, L\'{a}z\'{a}r, Papp, and S\'{a}ri]{Hartman2011}
Hartman, J. D., Bakos, G.~\'{A}., Torres, G., et al. 2011c, ApJ, 742, 59

\bibitem[Hellier et~al.(2009)Hellier, Anderson, Gillon, Lister, Maxted, Queloz, Smalley, Triaud, West, Wilson, Alsubai, Bentley, Cameron, Hebb, Horne, Irwin, Kane, Mayor, Pepe, Pollacco, Skillen, Udry, Wheatley, Christian, Enoch, Haswell, Joshi, Norton, Parley, Ryans, Street, and Todd]{Hellier2009}
Hellier, C., D.~R. Anderson, M.~Gillon, et al. ApJ, 690, L89

\bibitem[Horch et~al.(2009)]{Horch2009}
Horch, E. P., Veillette, D. R., Galle, R. B., et al. 2009, AK, 137, 6

\bibitem[Howard et~al.(2012)Howard, Bakos, Hartman, Torres, Shporer, Mazeh, Kov\'{a}cs, Latham, Noyes, Fischer, Johnson, Marcy, Esquerdo, B\'{e}ky, Butler, Sasselov, Stefanik, Perumpilly, L\'{a}z\'{a}r, Papp, and S\'{a}ri]{Howard2012}
Howard, A. W., Bakos, G.~\'{A}., Hartman, J., et al. 2012, ApJ, 749, 134

\bibitem[Husser et~al.(2013)Husser, {Wende-von Berg}, Dreizler, Homeier, Reiners, Barman, and Hauschildt]{Husser2013}
Husser, T.-O., Wende-von Berg, S., Dreizler, S., et al. 2013, A\&A, 553, A6

\bibitem[Jackson and Jeffries(2013)]{Jackson2013}
Jackson, R.J., \& Jeffries, R.D. 2008, MRNAS, 431, 2

\bibitem[Johns-Krull et~al.(2008)Johns-krull, Mccullough, Burke, Valenti, Janes, Heasley, Prato, Bissinger, Fleenor, Foote, Gary, and Howell]{Johns-Krull2008}
Johns-Krull, C. M., McCullough, P. R., Burke, C. J., et al. 2008, ApJ, 677, 657

\bibitem[Johnson et~al.(2009)Johnson, Winn, Cabrera, and Carter]{Johnson2009}
Johnson, J. A., Winn, J. N., Cabrera, N. E., \& Carter, J. A. 2009b, ApJL, 692, L100

\bibitem[Johnson et~al.(2011)Johnson, Winn, Bakos, Hartman, Morton, Torres, Kov\'{a}cs, Latham, Noyes, Sato, Esquerdo, Fischer, Marcy, Howard, Buchhave, Fűr\'{e}sz, Quinn, B\'{e}ky, Sasselov, Stefanik, L\'{a}z\'{a}r, Papp, and S\'{a}ri]{Johnson2011}
Johnson, J. A., Winn, J. N., Bakos, G.~\'{A}., et al. 2011, ApJ, 735, 24

\bibitem[Joshi et~al.(2009)Joshi, Pollacco, Cameron, Skillen, Simpson, Steele, Street, Stempels, Christian, Hebb, Bouchy, Gibson, H\'{e}brard, Keenan, Loeillet, Meaburn, Moutou, Smalley, Todd, West, Anderson, Bentley, Enoch, Haswell, Hellier, Horne, Irwin, Lister, McDonald, Maxted, Mayor, Norton, Parley, Perrier, Pont, Queloz, Ryans, Smith, Udry, Wheatley, and Wilson]{Joshi2009}
Joshi, Y. C., Pollacco, D., Collier Cameron, A., et al. 2009, MNRAS, 392, 1532

\bibitem[Kipping et~al.(2010)Kipping, Bakos, Hartman, Torres, Shporer, Latham, Kov\'{a}cs, Noyes, Howard, Fischer, Johnson, Marcy, B\'{e}ky, Perumpilly, Esquerdo, Sasselov, Stefanik, L\'{a}z\'{a}r, Papp, and S\'{a}ri]{Kipping2010}
Kipping, D. M., Bakos, G.~\'{A}., Hartman, J., et al. 2010, ApJ, 725, 2017

\bibitem[Kipping et~al.(2011)Kipping, Hartman, Bakos, Torres, Latham, Bayliss, Kiss, Sato, B\'{e}ky, Kov\'{a}cs, Quinn, Buchhave, Andersen, Marcy, Howard, Fischer, Johnson, Noyes, Sasselov, Stefanik, L\'{a}z\'{a}r, Papp, S\'{a}ri, and Fűr\'{e}sz]{Kipping2011}
Kipping, D. M., Hartman, J., Bakos, G.~\'{A}., et al. 2011, AJ, 142, 95

\bibitem[Kley \& Nelson(2012)]{Kley2012}
Kley, W. \& Nelson, R.P. 2012, ARA\&A, 50, 211-249

\bibitem[Knutson et~al.(2010)Knutson, Howard, and Isaacson]{Knutson2010}
Knutson, H. A., Howard, A. W., \& Isaacson, H. 2010, ApJ, 720, 1569

\bibitem[Knutson et~al.(2014)Knutson, Fulton, Montet]{Knutson2014}
Knutson, H.~A., B.~J. Fulton, B.~T. Montet et al. 2014, ApJ, 785, 126

\bibitem[Kolbl et~al.(2015)Kolbl, Marcy, Isaacson, and Howard]{Kolbl2015}
Kolbl, R., G.~W. Marcy, H. Isaacson, et al.  2015, AJ, 149, 18 

\bibitem[Kovacs et~al.(2007)Kovacs, Bakos, Torres, Sozzetti, Latham, Noyes, Butler, Marcy, Fischer, Fernandez, Esquerdo, Sasselov, Stefanik, Pal, Lazar, Papp, and Sari]{Kovacs2007}
Kovacs, G., Bakos, G.~\'{A}., Torres, G., et al. 2007, ApJL, 670, L41

\bibitem[Kovacs et~al.(2010)Kovacs,  Hartman, Torres, Noyes, Latham, Howard, Fischer, Johnson, Marcy, Isaacson, Sasselov, Stefanik, Esquerdo, Fernandez, L\'{a}z\'{a}r, Papp, and S\'{a}ri]{Kovacs2010}
Kovacs, G., Bakos, G.~\'{A}., Hartman, J. D., et al. 2010, ApJ, 724, 866

\bibitem[Kraus et~al.(2012)Kraus, Ireland, Hillenbrand, and Martinache]{Kraus2012}
Kraus, A.~L., M.~J. Ireland, L.~A. Hillenbrand, et al. 2012, ApJ, 745, 19

\bibitem[Lai et~al.(2011)]{Lai2011}
Lai, D., Foucart, F., \& Lin, D. N. 2011, MNRAS, 412, 4

\bibitem[Latham et~al.(2009)Latham, Bakos, Torres, Stefanik, Noyes, Kov\'{a}cs, P\'{a}l, Marcy, Fischer, Butler, Sipőcz, Sasselov, Esquerdo, Vogt, Hartman, Kov\'{a}cs, L\'{a}z\'{a}r, Papp, and S\'{a}ri]{Latham2009}
Latham, D.~W. , G.~\'{A}. Bakos, G. Torres, et al.  2009, ApJ, 704, 2

\bibitem[Law et~al.(2014)]{Law2014}
Law, N. M., Morton, T., Baranec, C., et al. 2014, ApJ, 791, 35

\bibitem[Lee et~al.(2014)]{Lee2014}
Lee, E.~J., Chiang, E., Ormel, C.W. 2014, ApJ, 797, 95

\bibitem[Lillo-Box et~al. (2012)]{Lillo-Box2012}
Lillo-Box, J, Barrado, D., Buoy, H. 2012, A\&A, 546, 10 

\bibitem[Lin et~al.(1996)]{Lin1996}
Lin, D. N. C., Bodenheimer, P., \& Richardson, D. C. 1996, Nature, 380, 606

\bibitem[Lockwood et al. (2014)]{Lockwood2014}
Lockwood, A.C., Johnson, J.A., Bender, C.F., et al. 2014, ApJ, 783, L29 

\bibitem[Malmberg et~al.(2007)Malmberg, Davies, and Chambers]{Malmberg2008}
Malmberg, D., Davies, M. B., \& Chambers, J. E. 2007, MNRAS, 377, L1

\bibitem[Mancini et~al.(2013)Mancini, Southworth, Ciceri, Fortney, Morley, Dittmann, Tregloan-Reed, Bruni, Barbieri, Evans, D’Ago, Nikolov, and Henning]{Mancini2013}
Mancini, L., Southworth, J., Ciceri, S., et al. 2013b, A\&A, 551, A11

\bibitem[Maness et~al.(2007)Maness, Marcy, Ford, Hauschildt, Shreve, Basri, Butler, Vogt, and Maness]{Maness2007}
Maness, H. L., Marcy, G. W., Ford, E. B., et al. 2007, PASP, 119, 90

\bibitem[Marois et~al.(2006)]{Marois2006}
Marois, C., Lafreniere, D., Doyon, R., et al. 2006, ApL, 641, 1

\bibitem[Mayer et~al.(2005)Mayer, Wadsley, Quinn, and Stadel]{Mayer2005}
Mayer, L., J. Wadsley, T. Quinn, et al. 2005, MNRAS, 363, 2

\bibitem[McCullough et~al.(2008)McCullough, Burke, Valenti, Long, Johns-Krull, Machalek, Janes, Taylor, Gregorio, Foote, Gary, Fleenor, Garcia-Melendo, and Vanmunster]{McCullough2008}
McCullough, P. R., Burke, C. J., Valenti, J. A., et al. 2008, arXiv:0805.2921

\bibitem[McLean et~al.(1998)]{McLean1998}
McLean, I. S., Becklin, E. E., Bendiksen, O., et al. 1998, Proc. SPIE 3354, 566

\bibitem[Miller et~al.(2010)Miller, {Collier Cameron}, Simpson, Pollacco, Enoch, Gibson, Queloz, Triaud, H\'{e}brard, Boisse, Moutou, and Skillen]{Miller2010}
Miller, G.~R.~M., A.~{Collier-Cameron}, E.~K. Simpson, et al. 2010, A\&A, 523, A52

\bibitem[Nagasawa et~al.(2008)Nagasawa, Ida, and Bessho]{Nagasawa2008}
Nagasawa, M., Ida, S., \& Bessho, T. 2008, ApJ, 678, 498

\bibitem[Naoz et~al.(2011)]{Naoz2011}
Naoz, S., Farr, W. M., Lithwick, Y., Rasio, F. A., \& Teyssandier,J. 2011, Nature, 473, 187

\bibitem[Ngo et~al.(2015)Ngo, Knutson, Hinkley, Crepp, Bechter, Batygin, Howard, Johnson, Morton, and Muirhead]{Ngo2015}
Ngo H., H.~A. Knutson, S. Hinkley, et al. 2015, ApJ, 800, 138  

\bibitem[Noyes et~al.(2008)Noyes, Bakos, and Torres]{Noyes2008}
Noyes, R. W., Bakos, G.~\'{A}., Torres, G., et al. 2008, ApJL, 673, L79

\bibitem[O'Donovan et~al.(2006)O'Donovan, Charbonneau, Mandushev, Dunham, Latham, Torres, Sozzetti, Brown, Trauger, Belmonte, Rabus, Almenara, Alonso, Deeg, Esquerdo, Falco, Hillenbrand, Roussanova, Stefanik, and Winn]{O'Donovan2006}
O'Donovan, F.~T., D. Charbonneau, G. Mandushev, et al.  2009, ApJ, 651, 1

\bibitem[P\'{a}l et~al.(2009)P\'{a}l, Bakos, Fernandez, Sipőcz, Torres, Latham, Kov\'{a}cs, Noyes, Marcy, Fischer, Butler, Sasselov, Esquerdo, Shporer, Mazeh, Stefanik, and Isaacson]{Pal2009}
P\'{a}l, A.~, Bakos, G.~\'{A}., Fernandez, J., et al. 2009, ApJ, 700, 783

\bibitem[P\'{a}l et~al.(2010)P\'{a}l, Bakos, Torres, Noyes, Fischer, Johnson, Henry, Butler, Marcy, Howard, Sipőcz, Latham, and Esquerdo]{Pal2010}
P\'{a}l,A.~, G.~\'{A}. Bakos, G. Torres, et al. 2010, MNRAS, 401, 4

\bibitem[Pichardo et~al.(2005)Pichardo, Sparke, Aguilar, and Astron]{Pichardo2005}
Pichardo, B., Sparke, L. S., \& Aguilar, L. A. 2005, MNRAS, 359, 521

\bibitem[Pollack et~al.(1996)]{Pollack1996}
Pollack, J.~, B., O.~Hubickyk, P. Bodenheimer, et al.  1996, Icarus, 124, 1. 

\bibitem[Pont et~al.(2008)Pont, Knutson, Gilliland, Moutou, and Charbonneau]{Pont2008}
Pont, F.~, H.~Knutson, R.~L. Gilliland, et al.  2008, MNRAS, 385, 1

\bibitem[Queloz et~al.(2010)Queloz, Anderson, {Collier Cameron}, Gillon, Hebb, Hellier, Maxted, Pepe, Pollacco, S\'{e}gransan, Smalley, Triaud, Udry, and West]{Queloz2010}
Queloz, D., Anderson, D. R., Collier Cameron, A., et al. 2010, A\&A, 517, L1

\bibitem[Raghavan et~al.(2010)Raghavan, McAlister, Henry, Latham, Marcy, Mason, Gies, White, and ten Brummelaar]{Raghavan2010}
Raghavan, D., McAlister, H. A., Henry, T. J., et al. 2010, ApJS, 190, 1

\bibitem[Sing et~al.(2011)Sing, Pont, Aigrain, Charbonneau, D\'{e}sert, Gibson, Gilliland, Hayek, Henry, Knutson, des Etangs, Mazeh, and Shporer]{Sing2011}
Sing, D.~K. , F.~Pont, S.~Aigrain, et al.  2011, MNRAS, 416, 2

\bibitem[Smalley et~al.(2011)Smalley, Anderson, Cameron, Hellier, Lendl, Maxted, Queloz, Triaud, West, Bentley, Enoch, Gillon, Lister, Pepe, Pollacco, Segransan, Smith, Southworth, Udry, Wheatley, Wood, and Bento]{Smalley2011}
Smalley, B., Anderson, D. R., Collier Cameron, A., et al. 2011, A\&A, 526, A130

\bibitem[Snellen et al.(2010)]{Snellen2010}
Snellen, I. A. G., de Kok, R. J., de Mooij, E. J. W., \& Albrecht, S. 2010, Nature, 465, 1049

\bibitem[Southworth et~al.(2013)Southworth, Mancini, Browne, Burgdorf, {Calchi Novati}, Dominik, Gerner, Hinse, Jorgensen, Kains, Ricci, Schafer, Schonebeck, Tregloan-Reed, Alsubai, Bozza, Chen, Dodds, Dreizler, Fang, Finet, Gu, Hardis, Harpsoe, Henning, Hundertmark, Jessen-Hansen, Kerins, Kjeldsen, Liebig, Lund, Lundkvist, Mathiasen, Nikolov, Penny, Proft, Rahvar, Sahu, Scarpetta, Skottfelt, Snodgrass, Surdej, and Wertz]{Southworth2013}
Southworth, J., Mancini, L., Browne, P., et al. 2013, MNRAS, 434, 1300

\bibitem[Southworth(2012)]{Southworth2012}
Southworth, J., Hinse, T. C., Dominik, M., et al. 2012b, MNRAS, 426, 1338

\bibitem[Sozzetti et~al.(2009)Sozzetti, Torres, Charbonneau, Winn, Korzennik, Holman, Latham, Laird, Fernandez, O'Donovan, Mandushev, Dunham, Everett, Esquerdo, Rabus, Belmonte, Deeg, Brown, Hidas, and Baliber]{Sozzetti2009}
Sozzetti, A., Torres, G., Charbonneau, D., et al. 2009, ApJ, 691, 1145

\bibitem[Spalding \& Batygin(2014)]{Spalding2014}
Spalding, C. \& Batygin, K. 2014, ApJ, 790, 1

\bibitem[Spalding et~al.(2014)]{Spalding2014b}
Spalding, C., Batygin, K., \& Adams, F. C. 2014. ApJL, 797, 2

\bibitem[Spalding \& Batygin(2015)]{Spalding2015}
Spalding, C. \& Batygin, K. 2015, ApJ, accepted

\bibitem[Spurzem et~al.(2009)]{Spurzem2009}
Spurzem, R., Giersz, M., Heggie, D.C., et al. 2009, ApJ, 697, 458

\bibitem[Street et~al.(2010)Street, Simpson, Barros, Pollacco, Joshi, Todd, {Collier Cameron}, Enoch, Parley, Stempels, Hebb, Triaud, Queloz, Segransan, Pepe, Udry, Lister, Depagne, West, Norton, Smalley, Hellier, Anderson,  Maxted, Bentley, Skillen, Gillon, Wheatley, Bento, Cathaway-Kjontvedt, and Christian]{Street2010}
Street, R. A., Simpson, E., Barros, S. C. C., et al. 2010, ApJ, 720, 337

\bibitem[Terquem \& Bertout(1993)]{Terquem1993}
Terquem, C. \& Bertout, C. 1993, A\&A, 274, 291

\bibitem[Teske et~al.(2015)]{Teske2015}
Teske, J.K., Everett, M.E., Hirsch, L, et al. 2015, accepted to AJ, arXiv:1508.06502

\bibitem[Torres et~al.(2010)Torres, Hartman, Noyes, Latham, Esquerdo, Fischer,  Johnson, Marcy, Butler, Sasselov, Fernandez, Stefanik, Isaacson, Howard,  Vogt, and Papp]{Torres2010}
Torres, G., Bakos, G. A., Hartman, J., et al. 2010, ApJ, 715, 458

\bibitem[Torres(2007)]{Torres2007}
Torres, G. 2007, ApJ, 671, 1

\bibitem[Torres et~al.(2012)Torres, Fischer, Sozzetti, Buchhave, Winn, Holman, and Carter]{Torres2012}
Torres, G., Fischer, D. A., Sozzetti, A., et al. 2012, ApJ, 757, 161

\bibitem[Triaud et~al.(2010)Triaud, Cameron, Queloz, Anderson, Hellier, Maxted, Mayor, Pepe, Pollacco, Smalley, West, and Wheatley]{Triaud2010}
Triaud, A. H. M. J., Collier Cameron, A., Queloz, D., et al. 2010, A\&A, 524, A25

\bibitem[Van Eylen et~al.(2012)Eylen, Kjeldsen, and Aerts]{Eylen2012}
Van Eylen, V., Kjeldsen, H., Christensen-Dalsgaard, J., \& Aerts, C. 2012, AN, 333, 1088

\bibitem[von Braun et~al.(2012)von Braun, Boyajian, Kane, Hebb, van Belle, Farrington, Ciardi, Knutson, ten Brummelaar, L\'{o}pez-Morales, McAlister, Schaefer, Ridgway, {Collier Cameron}, Goldfinger, Turner, Sturmann, and Sturmann]{VonBraun2012}
von Braun, K., Boyajian, T. S., Kane, S. R., et al. 2012, ApJ, 753, 171

\bibitem[Wang et~al(2014a)]{Wang2014a}
Wang, J., Xie, J.-W., Fischer, D. A., et al. 2014, ApJ, 783, 4

\bibitem[Wang et~al(2014b)]{Wang2014b}
Wang, J., Fischer, D. A., Xie, J.-W. et al. 2014, ApJ, 791, 111

\bibitem[Wang et~al.(2015)]{Wang2015}
Wang, J., Fischer, D.A., Xie, J.-W. et al. 2015, ApJ accepted, arXiv:1510.01964

\bibitem[Welsh et~al.(2012)]{Welsh2012}
Welsh, W. F., Orosz, J. A., Carter, et al. 2012, Nature, 481, 475

\bibitem[Wilson et~al.(2008)Wilson, Gillon, Hellier, Maxted, Pepe, Queloz, Anderson, Cameron, Smalley, Lister, Bentley, Blecha, Christian, Enoch, Haswell, Hebb, Horne, Irwin, Joshi, Kane, Marmier, Mayor, Parley, Pollacco, Pont, Ryans, Segransan, Skillen, Street, Udry, West, and Wheatley]{Wilson2008}
Wilson, D. M., Gillon, M., Hellier, C., et al. 2008, ApJL, 675, L113

\bibitem[Winn et~al.(2008)Winn, Holman, Torres, McCullough, Johns‐Krull, Latham, Shporer, Mazeh, Garcia‐Melendo, Foote, Esquerdo, and Everett]{Winn2008}
Winn, J. N., Holman, M. J., Torres, G., et al. 2008a, ApJ, 683, 1076

\bibitem[Winn et~al.(2010)Winn, Fabrycky, Albrecht, and Johnson]{Winn2010}
Winn, J. N., Fabrycky, D., Albrecht, S., \& Johnson, J. A. 2010a, ApJL, 718, L145

\bibitem[Winn et~al.(2011)Winn, Howard, Johnson, Marcy, Isaacson, Shporer, Bakos, Hartman, Holman, Albrecht, Crepp, and Morton]{Winn2011}
Winn, J. N., Howard, A. W., Johnson, J. A., et al. 2011, AJ, 141, 63

\bibitem[Woellert et~al.(2015)]{Woellert2015a}
Woellert, M., Brandner, W., Bergfors, C., et al. 2015, A\&A, 575, 23

\bibitem[Woellert \& Brandner(2015)]{Woellert2015}
Woellert, M. \& Brandner, W. 2015, A\&A, 579, 129

\bibitem[Wright et~al.(2012)Wright, Marcy, Howard, Johnson, Morton, and Fischer]{Wright2012}
Wright, J.~T., G.~W. Marcy, A.~W. Howard, et al. 2012, ApJ, 753, 2

\bibitem[Wu \& Lithwick(2010)]{Wu2010}
Wu, Y., \& Lithwick, Y. 2010, ApJ, 735, 109

\bibitem[Wu \& Murray(2003)]{Wu2003}
Wu, Y., \& Murray, N. 2003, ApJ, 589, 605

\bibitem[Zheng et~al.(2015)]{Zheng2015}
Zheng et al. 2015, MNRAS, 453, 2759

\bibitem[Zucker \& Mazeh(1994)]{Zucker1994}
Zucker, S., \& Mazeh, T. 1994, ApJ, 420, 806

\end{thebibliography}
\end{document}